\pdfoutput=1
\documentclass[cernpreprint, atlasdraft=false, texlive=2025, UKenglish, texmf, orcidlogo]{atlasdoc}
\usepackage{atlaspackage}
\usepackage{atlasbiblatex}

\usepackage{atlasphysics}
\graphicspath{{logos/}{figures/}}
\usepackage{comment}
\usepackage{ANA-HIGP-2024-30-PAPER-defs}

\AtlasTitle{Search for  Higgs boson production in association with another Higgs or an exotic scalar boson in the $b\bar{b}W^{+}W^{-}$ single lepton final state at $\sqrt{s}=13$ TeV with the ATLAS experiment}

\AtlasVersion{2.4}

\AtlasAbstract{%
A search for Higgs boson pair production where one Higgs boson decays
via $H \to b \bar{b}$ and the other Higgs boson decays via $H \to WW^*
\to \ell \nu q \bar{q}\quad (\ell=e,\mu)$ is
presented. This search uses the complete Run 2 data sample recorded by the ATLAS detector
at the LHC from 2015 to 2018 at a centre-of-mass energy of 13~TeV,
corresponding to an integrated luminosity of $140$~fb$^{-1}$.
The analysis looks for non-resonant Higgs boson pair
production assuming Standard Model dynamics.
No evidence for Higgs-boson pair production is observed and an upper limit at 95\% confidence level is set at 35 times the Standard Model prediction.
The same selection is used to target Beyond the Standard Model theories predicting the resonant production of the heavy scalars,
$X$
and $S$, with $m_X > m_S > m_H$, undergoing the process $X \to SH, S \to W^+W^-, H \to b \bar{b}$.
Limits are set on the resonant $X \to SH$ production for several combinations of the
two scalar's masses within the $m_X$ mass range $350-3000$\,GeV and $m_S$ mass range $170-2500$\,GeV.
}

\AtlasRefCode{HIGP-2024-30}

\PreprintIdNumber{CERN-EP-2026-255}

\AtlasJournal{JHEP}
\hypersetup{pdftitle={ATLAS document},pdfauthor={The ATLAS Collaboration}}

\begin{document}

\maketitle

\section{Introduction}
\label{sec:intro}

\begin{figure}[!h]
\begin{center}
\centering
\subfloat[]{
\begin{minipage}[c][110pt]{140pt}
\includegraphics[width=\textwidth]{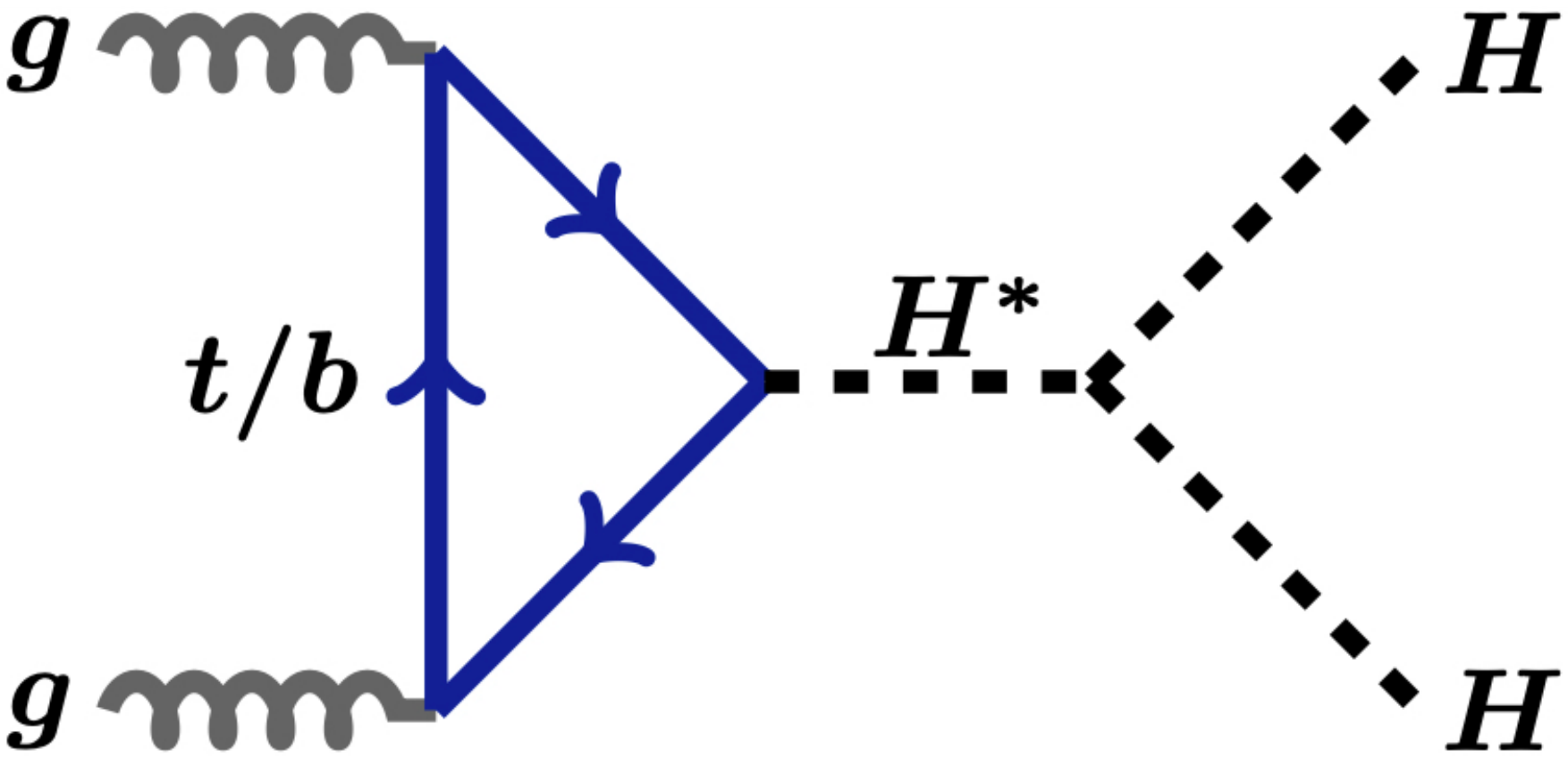}
\end{minipage}
}
\subfloat[]{
\begin{minipage}[c][110pt]{140pt}
\includegraphics[width=\textwidth]{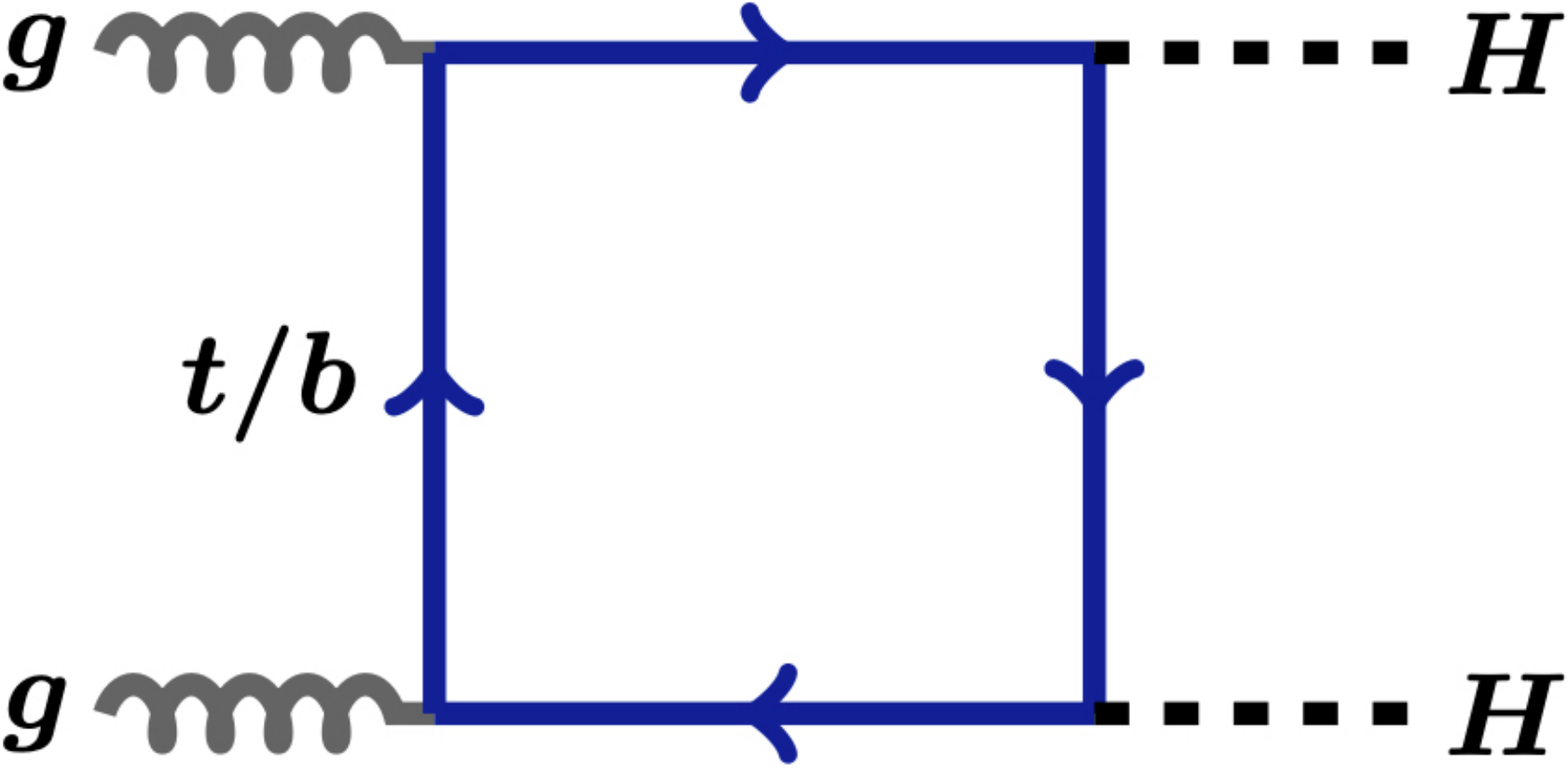}
\end{minipage}
}
\subfloat[]{
\begin{minipage}[c]{140pt}
\includegraphics[width=\textwidth]{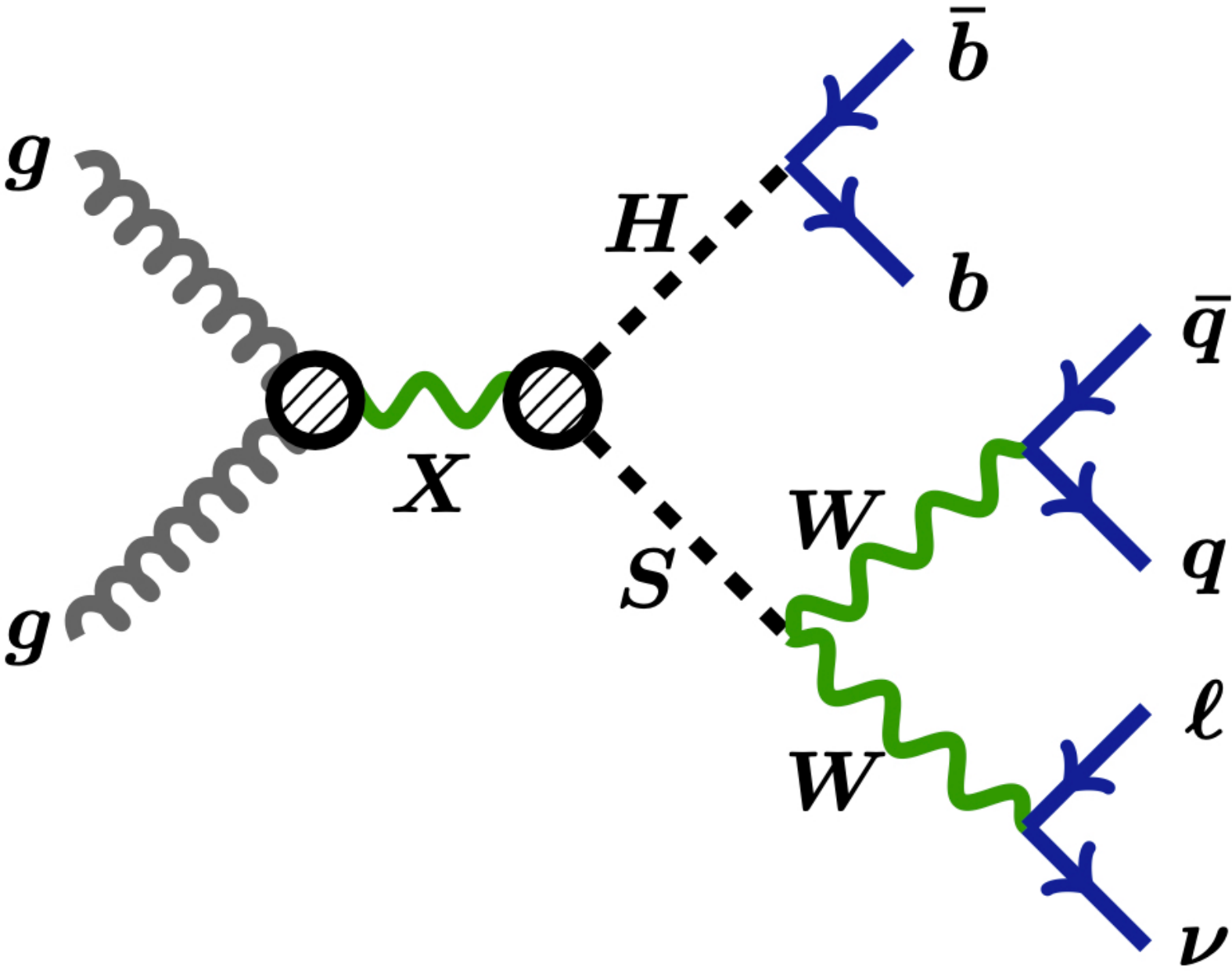}
\end{minipage}
}
\caption{
Leading-order Feynman diagrams for the production of non-resonant Higgs boson pairs and BSM $X \to SH$~:
(a) the $HH$ production diagram contains a Higgs
boson self-coupling vertex;  (b) the $HH$ production diagram contains the Higgs--fermion Yukawa
interaction; (c) the diagram presents the $SH$ production process. The $H^*$ is the off-shell Higgs boson
mediator, $X$ indicates the scalar mediator producing the $SH$ pair and $S$ is a heavy scalar decaying to a $WW$ pair. }  \label{fig:hh_production}
\end{center}
\end{figure}

The Higgs boson ($H$) is an essential part of the Standard Model (SM) and it
has a crucial role  in the electroweak symmetry breaking
mechanism~\cite{PhysRevLett.13.321,HIGGS1964132,PhysRevLett.13.508,PhysRevLett.13.585,PhysRev.145.1156,PhysRev.155.1554}.
In this mechanism, an SU(2) doublet bosonic scalar field is subject to a potential energy term whose shape allows the doublet field to acquire a vacuum expectation value. This breaks the SU(2) symmetry thereby producing the massive Higgs boson as well as the mass of all fermions and the $W$ and $Z$ bosons.
The potential energy term is the experimentally least probed component of the SM Lagrangian and thus requires further precision measurements.
The Higgs boson potential in the SM has only two parameters, which can be taken
to be the Fermi coupling constant $G_F$ and the Higgs boson mass $m_H$.
The shape of the Higgs potential can be tested directly through
the measurement of the Higgs self-coupling. Higgs-boson pair production in proton--proton $(pp)$ collisions proceeds at leading order through the self-interaction diagram and the quark-loop contributions via the top--Higgs Yukawa coupling, as illustrated in \Fig{\ref{fig:hh_production}(a)}  and \Fig{\ref{fig:hh_production}(b)}, respectively. The SM cross-section for this process is extremely small, it is dominated by gluon--gluon fusion with a cross-section of $31.1^{+1.9}_{-7.1}$~fb (scale + $m_{\textrm{top}}$) $\pm$ 0.9 (PDF+$\alpha_s$)~fb~\cite{deFlorian:2016spz,Grazzini2018,Baglio:2020wgt} at 13~\TeV.
Physics beyond the SM (BSM) can manifest itself in a scenario where a SM Higgs boson $H$ is produced together with a new heavy scalar $S$ through the decay of a heavy scalar resonance $X$ produced in gluon--gluon fusion. \Fig{\ref{fig:hh_production}}(c) shows  a schematic diagram of this process, with subsequent decays $S \to WW$  and $H \to b \bar{b}$.
This BSM scenario is explored in  several theoretical models~\cite{Abouabid_2022}, e.g. the Two Higgs Doublet Models (2HDM) with additional singlet~\cite{Keus:2017ioh} or when extending the Standard Model with two additional singlets~\cite{Ahriche:2013vqa}.
These models can modify the Higgs energy potential to produce the observed matter/anti-matter asymmetry in the universe at the Electroweak symmetry breaking epoch~\cite{GiChol2026}.
Given the low expected yield for SM $HH$ non-resonant production,
it is of great importance to understand the sensitivity for the observation of the Higgs boson pair production in all possible decay channels, including $b \bar{b}WW^{\ast}$, which could give a valuable contribution when combined with the most sensitive channels and will improve projections for future high-luminosity
and high-energy $pp$ colliders. The $b \bar{b}WW^{\ast}$ channel,  while having the second highest branching fraction after the 4$b$ final state, is challenging due to the large
background from top-quark pair production $(t\bar{t})$, which features the same final state.
Furthermore, this final state is especially important for searches for a combined production of the Higgs boson with a new Higgs-like heavy scalar resonance, $S$, which is expected to have a branching fraction to the $WW$ final state higher than 55\% for $m_S > 2m_W$~\cite{Abouabid_2022}.
The minimal considered mass of the $S$ scalar is 170~\gev, which is high enough to produce two on-shell $W$ bosons.

In this paper a search for non-resonant Higgs-boson pair production in the $HH \to b\bar{b} W^+ W^-$ decay channel is presented. The $WW$ system decays into  $\ell\nu q\bar{q}^\prime$ (where $\ell$ is either an electron, a muon or a $\tau$-lepton decaying fully leptonically).  In addition, resonant production through the chain $pp \to X \to SH \to W^+W^- b \bar{b}$ is explored.
The heavy scalars, $S$ and $X$, are  assumed to have a narrow width (10~\MeV). The sum of the masses of the $S$ scalar and the Higgs boson is chosen to be smaller than the mass of $X$ to have a non-suppressed, on-shell $S$ and $H$ production. In both searches the SM $HH$ production is neglected and not included among the background processes.

The ATLAS and CMS collaborations have published search results for non-resonant $HH$ production in several decay channels using the complete data sample collected during LHC \RunTwo (from 2015 to 2018)~\cite{HDBS-2018-33,HDBS-2018-34,HDBS-2018-40,HDBS-2019-02,HDBS-2019-04,HDBS-2019-29,HDBS-2022-02, HDBS-2022-03,CMS-HIG-19-018,
CMS-HIG-20-004, CMS-HIG-20-005, CMS-HIG-20-010, CMS-HIG-21-002, CMS-HIG-21-005, CMS-B2G-22-003}
and using partial data sample collected during LHC \RunThree (from 2022 to 2024)~\cite{HIGP-2024-37,HIGP-2024-37,HIGP-2025-10,HIGP-2024-34,CMS-HIG-24-010,CMS-HIG-25-018}.
Furthermore, the two collaborations have  published statistical combinations of these double-Higgs boson search analyses~\cite{HDBS-2023-17,HDBS-2021-18,CMS-HIG-23-006}, using integrated luminosities of up to 140~fb$^{-1}$.
Previous measurements in the $HH \to b\bar{b}W^+W^- \to b \bar{b} \ell \nu q \bar{q}^\prime$ channel were performed by both ATLAS and CMS collaborations with the Run~2 data samples~\cite{HIGG-2016-27,CMS-HIG-21-005,CMS-B2G-18-008}. The ATLAS analysis used a data sample of 36~fb$^{-1}$ reaching an upper limit of 300 times the SM expected cross-section at 95\% confidence level (CL).
The CMS Collaboration combined analyses targeting the $b\bar{b}W^+W^-$ channel using both the single-lepton and the dilepton final states, while a dilepton final-state analysis of data taken during Run~3 (in 2022 and 2023) was presented in Ref.~\cite{CMS-HIG-25-018}.
The single-lepton channel was excluded at 28 times the SM cross-section at 95\% CL~\cite{CMS-HIG-21-005} using 138~fb$^{-1}$ of Run~2 data.

This work focuses on the single-lepton channel while the dilepton channel was previously published in Ref.~\cite{HDBS-2018-33} and combined with other dilepton modes in Ref.~\cite{HDBS-2019-02}.
The data analysed here were recorded by the ATLAS detector at the LHC in $pp$ collisions at $\sqrt{s}=13$~\TeV and correspond to an integrated luminosity of 140~fb$^{-1}$~\cite{DAPR-2021-01}.
The analysis requires two distinct $b$-jets, emerging from the  $H\rightarrow b \bar{b}$ decay in the signal case.
This ``resolved'' analysis explicitly excludes events where this decay is reconstructed as a wide single jet with radius $R = 1.0$ due to a large boost of the $b$-quark pair (often referred to as the ``boosted'' analysis). This topology will be studied in a future paper.

In addition to the dominant $t \bar{t}$ background, there are small
background contributions from $W$ bosons produced in association with jets ($W$+jets) and  multijet events in which a jet is
misidentified as a lepton.
The analysis  strategy is initially optimised without looking at regions with high expected signal contributions.


\section{The ATLAS detector}
\label{sec:detector}
%

\newcommand{\AtlasCoordFootnote}{
ATLAS uses a right-handed coordinate system with its origin at the nominal interaction point (IP)
in the centre of the detector and the \(z\)-axis along the beam pipe.
The \(x\)-axis points from the IP to the centre of the LHC ring,
and the \(y\)-axis points upwards.
Polar coordinates \((r,\phi)\) are used in the transverse plane,
\(\phi\) being the azimuthal angle around the \(z\)-axis.
The pseudorapidity is defined in terms of the polar angle \(\theta\) as \(\eta = -\ln \tan(\theta/2)\) and is equal to the rapidity
$ y = \frac{1}{2} \ln \left( \frac{E + p_z}{E - p_z} \right) $ in the relativistic limit.
Angular distance is measured in units of \(\Delta R \equiv \sqrt{(\Delta y)^{2} + (\Delta\phi)^{2}}\).}

The ATLAS detector~\cite{PERF-2007-01} at the LHC covers nearly the entire solid angle around the collision point.\footnote{\AtlasCoordFootnote}
It consists of an inner tracking detector surrounded by a thin superconducting solenoid, electromagnetic and hadronic calorimeters,
and a muon spectrometer incorporating three large superconducting air-core toroidal magnets.

The inner-detector system (ID) is immersed in a \qty{2}{\tesla} axial magnetic field
and provides charged-particle tracking in the range \(|\eta| < 2.5\).
The high-granularity silicon Pixel Detector covers the vertex region and typically provides four measurements per track,
the first hit generally being in the insertable B-layer installed before Run~2~\cite{ATLAS-TDR-19,PIX-2018-001}.
It is followed by the Semiconductor Tracker (SCT), which usually provides eight measurements per track.
These silicon detectors are complemented by the Transition Radiation Tracker (TRT),
which enables radially extended track reconstruction up to \(|\eta| = 2.0\).
The TRT also provides electron identification information
based on the fraction of hits (typically 30 in total) above a higher energy-deposit threshold corresponding to transition radiation.

The calorimeter system covers the pseudorapidity range \(|\eta| < 4.9\).
Within the region \(|\eta|< 3.2\), electromagnetic calorimetry is provided by barrel and
endcap high-granularity lead/liquid-argon (LAr) calorimeters,
with an additional thin LAr presampler covering \(|\eta| < 1.8\)
to correct for energy loss in material upstream of the calorimeters.
Hadronic calorimetry is provided by the steel/scintillator-tile calorimeter,
segmented into three barrel structures within \(|\eta| < 1.7\), and two copper/LAr hadronic endcap calorimeters.
The solid angle coverage is completed with forward copper/LAr and tungsten/LAr calorimeter modules
optimised for electromagnetic and hadronic energy measurements respectively.

The muon spectrometer (MS) comprises separate trigger and
high-precision tracking chambers measuring the deflection of muons in a magnetic field generated by the superconducting air-core toroidal magnets.
The field integral of the toroids ranges between \num{2.0} and \qty{6.0}~{ Tm}
across most of the detector.
Three layers of precision chambers, each consisting of layers of monitored drift tubes, cover the region \(|\eta| < 2.7\),
complemented by cathode-strip chambers in the forward region, where the background is highest.
The muon trigger system covers the range \(|\eta| < 2.4\) with resistive-plate chambers in the barrel, and thin-gap chambers in the endcap regions.

The luminosity is measured mainly by the LUCID--2~\cite{LUCID2} detector that records Cherenkov light produced in the quartz windows of photomultipliers located close to the beampipe.
Events are selected by the first-level trigger system implemented in custom hardware,
followed by selections made by algorithms implemented in software in the high-level trigger~\cite{TRIG-2016-01}.
The first-level trigger accepts events from the \qty{40}{\MHz} bunch crossings at a rate below \qty{100}{\kHz},
which the high-level trigger further reduces to record complete events to disk at about \qty{1.25}{\kHz}.
A software suite~\cite{SOFT-2022-02} is used in data simulation, in the reconstruction
and analysis of real and simulated data, in detector operations, and in the trigger and data acquisition
systems of the experiment.


%

%
\section{Data and simulated-event samples}
\label{sec:samples}

This search uses $pp$ collision data recorded at a centre-of-mass energy of $\sqrt{s}=13$~\TeV with the ATLAS detector at the LHC between 2015 and 2018,
corresponding to an integrated luminosity of 140.1$\pm1.2$\,fb$^{-1}$,
and satisfying data quality requirements for all sub-detectors~\cite{DAPR-2021-01,DAPR-2018-01}.
Samples of simulated signal and background events are used to
design the event selection and estimate the signal acceptance and
the background yields from various SM processes.
Table~\ref{tab:samples} shows the configurations for all Monte Carlo (MC) samples.
Samples used to estimate systematic uncertainties are indicated in parentheses.

\begin{table}[!htb]
\caption{The configurations used for event generation of signal and background processes.
The samples used to estimate the systematic uncertainties are shown in parentheses and grey text colour.
The column labelled ``Parton shower'' refers to the generator used for parton shower (PS), hadronisation and underlying events.
Tune refers to the underlying-event tune of the PS generator. Further details are given in the text.}
\label{tab:samples}
\begin{center}
\begin{footnotesize}
\begin{tabular}{lllll}
\toprule
Process                                    & Generator                & Parton shower                       & PDF            & Tune \\
\midrule
ggF~$HH$ signal                            & \textsc{Powheg-Box}      & \textsc{Pythia8}                    & PDF4LHC        & \textsc{A14}  \\
& {\color{gray}(\textsc{Powheg-Box})}    & {\color{gray}(\textsc{Herwig7})}                  & {\color{gray}(MMHT2014LO)}  & {\color{gray}(H7.1-Default)} \\
VBF~$HH$ signal                            & \textsc{MadGraph5}       & \textsc{Pythia8}                    & NNPDF3.0 NLO   & \textsc{A14}  \\
& {\color{gray}(\textsc{MadGraph5})}     & {\color{gray}(\textsc{Herwig7})}                  & {\color{gray}(NNPDF3.0 NLO)} & {\color{gray}(H7 default)}  \\
$X \to SH$ signal                          & \textsc{Pythia8}         & \textsc{Pythia8}                    & NNPDF2.3 LO    & \textsc{A14}  \\
\midrule
$t\bar{t}$                                 & \textsc{Powheg-Box}      & \textsc{Pythia8}                    & NNPDF3.0 NLO   & \textsc{A14}  \\
& {\color{gray}(\textsc{Madgraph5\_aMC@NLO})}      & {\color{gray}(\textsc{Pythia8})}                  & {\color{gray}(NNPDF3.0 NLO)} & {\color{gray}(A14)}   \\
& {\color{gray}(\textsc{Powheg-Box})}    & {\color{gray}(\textsc{Herwig7})}                  & {\color{gray}(NNPDF3.0 NLO)} & {\color{gray}(H7 default)}   \\
& {\color{gray}(\textsc{Powheg-Box})}    & {\color{gray}(\textsc{Pythia8} ($\textsc{Hdamp}=3.0\,m_t$))} & {\color{gray}(NNPDF3.0 NLO)} & {\color{gray}(A14)}   \\
$Wt$                  & \textsc{Powheg-Box}      & \textsc{Pythia8} (DR-scheme)        & NNPDF3.0 NLO   & \textsc{A14}  \\
& {\color{gray}(\textsc{Powheg-Box})}    & {\color{gray}(\textsc{Pythia8} (DS-scheme))}      & {\color{gray}(NNPDF3.0 NLO)} & {\color{gray}(A14)}   \\
& {\color{gray}(\textsc{Powheg-Box})}    & {\color{gray}(\textsc{Herwig7})}                  & {\color{gray}(NNPDF3.0 NLO)} & {\color{gray}(H7 default)}   \\
& {\color{gray}(\textsc{Madgraph5\_aMC@NLO})}      & {\color{gray}(\textsc{Pythia8})}                  & {\color{gray}(NNPDF3.0 NLO)} & {\color{gray}(A14)}   \\
single-top ($s-$ and $t-$channel)          & \textsc{Powheg-Box}      & \textsc{Pythia8}                    & NNPDF3.0 NLO   & \textsc{A14}  \\
& {\color{gray}(\textsc{Powheg-Box})}    & {\color{gray}(\textsc{Herwig7})}                  & {\color{gray}(NNPDF3.0 NLO)} & {\color{gray}(H7 default)}  \\
& {\color{gray}(\textsc{Madgraph5\_aMC@NLO})}      & {\color{gray}(\textsc{Pythia8})}                  & {\color{gray}(NNPDF3.0 NLO)} & {\color{gray}(A14)}  \\
$t\bar{t}W/Z$                              & {\textsc{Madgraph5\_aMC@NLO}}        & \textsc{Pythia8}                    & NNPDF3.0 NLO   & \textsc{A14}  \\
$W(\to \ell \nu)$+jets                         & \textsc{Sherpa 2.2.1}    & \textsc{Sherpa}                     & NNPDF3.0 NNLO   & Sherpa default \\
$Z(\to \ell^+\ell^-/\nu\bar{\nu})$+jets          & \textsc{Sherpa 2.2.1}    & \textsc{Sherpa}                     & NNPDF3.0 NNLO   & Sherpa default \\
Diboson ($VV$)                             & \textsc{Sherpa 2.2.1}    & \textsc{Sherpa}                     & NNPDF3.0 NNLO   & Sherpa default \\
Single-Higgs ggF                & \textsc{Powheg-Box}      & \textsc{Pythia8}                    & CT10   & AZNLO \\
Single-Higgs VBF                & \textsc{Powheg-Box}      & \textsc{Pythia8}                    & NNPDF3.0 NLO   & AZNLO \\
Single-Higgs $t\bar{t}H$                       & \textsc{Powheg-Box}      & \textsc{Pythia8}                    & NNPDF3.0 NLO   & \textsc{A14}  \\
Single-Higgs $ZH$,$WH$                   & \textsc{Powheg MiNLO}    & \textsc{Pythia8}                    & NNPDF3.0 NLO   & AZNLO \\
Multijet                                   & \textsc{Pythia8}         & \textsc{Pythia8}                    & NNPDF2.3 LO    & \textsc{A14} \\
\bottomrule
\end{tabular}
\end{footnotesize}
\end{center}
\end{table}

The  $HH$ gluon--gluon fusion (ggF) production signal samples were
simulated with the \textsc{Powheg Box} v2 generator~
\cite{Alioli:2010xd,Nason:2004rx,Frixione:2007vw,Heinrich_2017} at next-to-leading-order (NLO),
including finite top-quark mass effects~\cite{PhysRevLett.117.012001} and PDF4LHC
parton density functions (PDF)~\cite{Butterworth:2015oua}.
\textsc{Pythia8} (v8.244)~\cite{Sjostrand:2014zea} was used with the \textsc{A14} tune~\cite{ATL-PHYS-PUB-2014-021} for showering and  hadronisation of the nominal samples,
while alternative samples were interfaced with \textsc{Herwig7} (v7.1.6)~\cite{Bahr:2008pv,Bellm:2015jjp} as parton shower algorithm,
with \textsc{H7.1-Default} tune and \textsc{MMHT2014LO} PDF~\cite{Harland-Lang:2014zoa}.
The samples were normalised using a
production cross-section of $31.1^{+2.1}_{-7}$ fb~\cite{PhysRevD.58.115012,PhysRevLett.117.012001,Baglio2019,PhysRevLett.111.201801,Shao2013,deFlorian2025,Grazzini2018,Baglio2019}
. All signal samples were simulated with
100\% of Higgs boson pairs decaying into $b\bar{b}WW^{\ast}$, and
the samples were then normalised assuming ${\mathcal{B}}(H \to WW^*) =
0.2137$ and ${\mathcal{B}}(H \to bb) = 0.5824$~\cite{deFlorian:2016spz}. Contributions from other $HH$ decay processes were found to be negligible and were not included in  the MC samples.
The vector boson fusion (VBF) sample was generated with \textsc{Madgraph}~\cite{Alwall:2011uj} at leading-order (LO) in QCD with NNPDF3.0 NLO PDF~\cite{Ball:2014uwa},
interfaced to \textsc{Pythia8} (v8.244) with the \textsc{A14} tune. The sample was normalised to a cross-section of $1.726\pm 2.1\%$~fb
~\cite{Dreyer:2020xaj,Dreyer:2018qbw}.

Signal samples for the $X \to SH$ process were generated via gluon--gluon fusion at leading order (LO) using \textsc{Pythia8}~\cite{Sjostrand:2007gs} with the NNPDF2.3~\cite{Ball:2012cx} LO PDF set and the \textsc{A14} tune for generation, showering and hadronisation. The decay widths of the $X$ ($\Gamma_X$) and the $S$ ($\Gamma_S$) bosons were set to 10~\MeV.
The $S$ scalar always decays to a $WW$ pair at the generator level, while $H$ decays into a pair of $b$-jets.
A set of signal samples were simulated using the mass values tabulated in \Fig{\ref{fig:shgroups}}.

Background samples for $W \to \ell \nu$/$Z\to \ell\ell$+jets, and diboson ($WW$, $WZ$, and $ZZ$) production were generated with \textsc{Sherpa} v2.2.1~\cite{Gleisberg:2008ta} using the NNPDF3.0 NNLO PDF set. The $W/Z$+jets samples were normalised to an NNLO cross-section using FEWZ~\cite{Gavin:2010az}, whereas the diboson samples were normalised to NLO~\cite{Campbell:2011bn}.

The $t\bar{t}$ background samples were simulated with
\textsc{Powheg-Box} v2~\cite{Frixione:2007vw} using the NNPDF3.0 NLO PDF set~\cite{Ball:2017nwa}.
\textsc{Pythia8} (v8.230) was used for showering and hadronisation,
with the \textsc{A14} set of tuned underlying-event parameters.
The mass of the top quark was set to
$m_{t} = 172.5~\GeV$. At least one top quark in the $t\bar{t}$ event was
required to decay into a final state with a lepton.
The \textsc{Hdamp} parameter,
used to regulate the high-\pt\ gluon emission in \textsc{Powheg}, was set
to 1.5 times $m_{t}$, giving good modelling of the high-\pt\
region~\cite{ATL-PHYS-PUB-2014-005,ATL-PHYS-PUB-2016-020}. Due to the small width of the Higgs $(\Gamma_H\sim 4~\MeV)$, $S$ and $X$ bosons, interference
between the $t\bar{t}$ background and the signal is extremely small and
neglected.  The $t\bar{t}$
cross-section, used to normalise the sample, is calculated to next-to-next-to-leading-order in QCD including resummation of soft gluon
contributions at next-to-next-to-leading-logarithm (NNLL) accuracy using
\textsc{Top++} 2.0~\cite{Czakon:2011xx}.
Alternative $t\bar{t}$ samples with semileptonic final states were produced using the \textsc{Madgraph5\_aMC@NLO} generator for matrix element (ME) calculation, \textsc{Herwig7} for PS modelling,
and different \textsc{Hdamp} parameter to evaluate modelling uncertainties related to the choice of the nominal generators and parameters.
The contribution from resonant $t\bar{t}$ production near threshold, recently observed by the ATLAS and CMS collaborations~\cite{TOPQ-2025-11,CMS-TOP-24-007}, is neglected.
Top-quark pair production  in association with $W$ and $Z$ bosons, $t\bar{t}W$ and $t\bar{t}Z$, were simulated with  \textsc{Madgraph5\_aMC@NLO} at next-to-leading-order using the NNPDF3.0 PDF set, interfaced with the \textsc{Pythia8} parton shower
and using the \textsc{A14} tune for the underlying event.

Single top-quark events in the $Wt$-, s- and t-channels were simulated using \textsc{Powheg-Box} v1~\cite{Alioli:2009je, Re:2010bp}.
The single top-quark (single-top) $t$-channel and $s$-channel processes were normalised using a cross-section calculated at NLO in QCD using the \textsc{Hathor v2.1}
cross-section calculator~\cite{Kant:2014oha, Aliev:2010zk}.
The $Wt$-channel,  instead,  was normalised using the NNLO approximate predictions
calculated following the prescriptions in Ref.~\cite{Kidonakis:2013zqa, Kidonakis:2010ux}.
An alternative $Wt$-channel sample was produced using the Diagram Subtraction (DS) scheme, instead of the Diagram Removal (DR) scheme~\cite{Frixione:2008yi},
and alternative samples were produced for all channels using \textsc{Madgraph5\_aMC@NLO} generator for ME calculation,
or \textsc{Herwig7} generator for PS modelling, to provide ME and PS variations of the nominal samples.

Single Higgs processes with a similar final state may contribute to the background due to the small cross-section of Higgs pair production processes. All Higgs production modes (ggF, VBF, $t\bar{t}H$, $WH$, $ZH$) were simulated with \textsc{Powheg-Box} framework at NLO accuracy interfaced to \textsc{Pythia8} for parton showering. All samples use the AZNLO tune~\cite{STDM-2012-23}, except for the $t\bar{t}tH$ sample that uses the \textsc{A14} tune. All relevant Higgs-boson decay modes were considered. The samples were normalised to the highest order available predictions~\cite{deFlorian:2016spz}.
In particular the ggF~$H$ sample, with inclusive Higgs-boson decays, was simulated using the CT10 PDF set~\cite{Lai:2010vv} and normalised to the N3LO (QCD)+NLO(EW) cross-section~\cite{deFlorian:2016spz}.

The $t\bar{t}H$ process was simulated in three split samples depending on the top decay: all-hadronic, semileptonic and di-leptonic, with the inclusive Higgs decays.
The $VH$ samples were simulated with up to two additional jets, using the \textsc{MiNLO} method~\cite{Hamilton:2012np} with NLO calculations.
VBF and $ZH$ samples include inclusive $H$ and $Z$ boson decays, while the $WH$ sample includes only $H \to b\bar{b}$ and $W \to \ell \nu$ decays.

Finally, dijet samples are used to assess the extrapolation uncertainty in the data-driven background estimate.
These samples were simulated at leading-order using \textsc{Pythia8} with NNPDF2.3 LO PDF set.
The \textsc{A14} tune was applied to model parton showers and the hadronisation process.

\textsc{EvtGen}~\cite{Lange:2001uf} was used to simulate the bottom and charm hadron decays for all samples, except for those  generated with \textsc{Sherpa}.
The generation of the simulated event samples included the effect of multiple $pp$ interactions per bunch crossing (in-time pile-up)
and the effect on the detector response of interactions from bunch crossings before or after the one
containing the hard interaction (out-of-time pile-up).
A reweighting was applied to the simulated samples to correct the pile-up profile
to match that observed in data.
The samples were processed through a \textsc{Geant}4-based detector
simulation~\cite{Agostinelli:2002hh, SOFT-2010-01} with the
standard ATLAS reconstruction software~\cite{ATL-SOFT-PUB-2021-001} used for collision data.


\section{Object Selection}

\label{sec:objects}
Vertices from $pp$ collisions are reconstructed if they have associated at least two ID tracks with $\pT > 0.5$~\gev~\cite{IDTR-2021-01}.
The vertex with the largest $\sum \pt^2$ is selected as the primary vertex.
In the present work an ``object'' is defined to be a reconstructed jet, electron, or muon.
Electrons are reconstructed by associating clusters in the electromagnetic calorimeter with tracks in the Inner Detector. They are identified using a likelihood based algorithm and
are required to satisfy the ``Medium'' or ``Tight'' selection criteria~\cite{EGAM-2018-01} depending on the online trigger selection,
to have $\pt > 5$~\gev and to be within $|\eta|<2.47$, excluding the transition region between the barrel and
endcaps in the LAr calorimeter ($1.37 < |\eta| < 1.52$).
In addition, electrons are required to be isolated satisfying the ``Loose'' isolation criteria~\cite{PERF-2017-01},
and to have tracks compatible with the primary vertex requiring that the transverse impact
parameter significance $|d_0|/\sigma(d_0)<5$ and the longitudinal impact parameter $|\Delta z_0 \sin\theta|<0.5$\,mm.

Muons are reconstructed from MS track segments matched to an ID track~\cite{PERF-2015-10,MUON-2018-03}. They are required to satisfy
the ``Medium'' identification criterion, and have $\pt > 5$~\gev and $|\eta|<2.7$.
Furthermore, muons are required to be isolated according to the ``TightTrackOnly'' isolation criteria~\cite{DAPR-2021-01},
and selections are applied on the impact parameter significance $|d_0|/\sigma(d_0)<3$, and on the longitudinal impact parameter $|\Delta z_0 \sin\theta|<0.5$\,mm.

Hadronically decaying $\tau$-leptons (\tauhadvis) are used to remove events that overlap with the analysis targeting $b\bar{b}\tau^+\tau^-$ final state, their reconstruction is detailed in Ref.~\cite{ATL-PHYS-PUB-2015-045}.
A recurrent neural network (RNN) classifier is used to identify the visible decay products of hadronic $\tau$-lepton decays
as described in Ref.~\cite{ATL-PHYS-PUB-2019-033}.
Events are vetoed if at least one \tauhadvis candidate with $\pt > 20$~\gev, $|\eta|<2.5$,
one or three tracks, unit charge, and satisfying the loose working point of the RNN identification algorithm is found.

Jets are reconstructed using the anti-$k_t$ algorithm~\cite{Cacciari:2008gp} as implemented in FastJet~\cite{Fastjet} with a jet radius parameter $R=0.4$.
The inputs to this algorithm are particle flow (PFlow) objects~\cite{PERF-2015-09}, which combine measurements from the ATLAS inner detector
and calorimeters~\cite{PERF-2014-07} to improve the jet energy resolution and increase the jet reconstruction efficiency, especially at low jet \pt.
Jet cleaning is applied to remove events with jets built from noisy calorimeter cells or non-collision backgrounds.
Additionally, jets with $20<\pt<60$~\gev and $|\eta|<2.4$ must satisfy the tight working point of a discriminant called ``Jet Vertex Tagger'' (JVT)~\cite{PERF-2014-03}, which suppresses pile-up jets.
Jets are further required to have $\pt>20$~\gev ~and $|\eta|<2.5$.

Large-$R$ jets are reconstructed from topological energy depositions using the anti-$k_t$ algorithm with a jet radius parameter of $R=1.0$.  These jets are used only to avoid the overlap with boosted topologies, as described below.
The jets are trimmed~\cite{Krohn:2009:Trimming, JETM-2018-06} to reduce pile-up contributions and soft radiation effects.
A Monte Carlo simulation is used to account for the energy loss due to soft out-of-cone radiation and semileptonic $b$-hadron decays~\cite{JETM-2018-02}.
Large-$R$ jets must satisfy $\pt > 150$~\GeV and $|\eta| < 2.0$~\cite{JETM-2024-03}.

Jets containing  $b$-hadrons ($b$-jets) are identified using the DL1r algorithm~\cite{FTAG-2019-07},
using a working point with 77\% efficiency in a sample of jets with \pT of at least 20~\GeV from semileptonic $t\bar{t}$ events~\cite{FTAG-2018-01}.
Jets that are not tagged as $b$-jets are collectively referred to as ``light jets''.

The missing transverse momentum (MET) with magnitude
$\met$ is calculated as the negative vectorial sum of the
transverse momenta of all calibrated and selected
leptons and jets, and a soft term from ID tracks
matched to the primary vertex but not to any other object~\cite{JETM-2020-03}.

To avoid double-counting, overlapping objects are removed from the analysis according to the
following procedure:
\begin{itemize}
\item muons sharing their track with an electron  are removed if they are calorimeter-tagged. Otherwise, the electron is removed;
\item jets overlapping with electrons within an angular distance $\Delta R = 0.2$ are removed,
otherwise if they overlap within an angular distance $\Delta R = 0.4$, electrons are removed;
\item light jets overlapping with muons within $\Delta R = 0.2$ and having less than three tracks or carrying less than 50\%  of
the muon $\pt$ are removed,
otherwise the muon is removed if it overlaps with a jet within $\Delta R = 0.4$;
\end{itemize}
Large-$R$ jets are not considered in the overlap removal procedure.


\section{Event Selection}
\label{sec:selection}

\subsection{Event selection and classification}
\label{sec:event_selection}
At lowest-order in QCD, the final-state contains one charged lepton, one
neutrino and four quarks, two of which are $b$-quarks. The analysis targets the corresponding
detector signature with
one charged lepton ($e$/$\mu$), large $\met$,
and four or more jets. Two of these jets are $b$-jets from the Higgs boson decay,
and two jets are light jets from the hadronic $W$ boson decay.
Extra jets may arise due to QCD radiation.

Events are selected using electron, muon, or MET triggers~\cite{TRIG-2018-01,TRIG-2018-05,TRIG-2019-01}. For electron-triggered events, the lowest lepton \pt threshold was 24~\gev in 2015 and 26~\gev during 2016–2018, while for muon-triggered events, it was 20~\gev and 26~\gev, respectively. For MET-triggered events, the applied MET thresholds were 80~\gev in 2015, 100~\gev in 2016, and 120~\gev from 2017 onward.
To achieve a high and uniform trigger efficiency, trigger-matched electrons (muons) are required to exceed the nominal threshold by 1~\gev (5\%). Similarly, the MET threshold is required to exceed the nominal trigger threshold by 10~\gev.

Events are required to have exactly two
$b$-tagged jets, which form the Higgs boson candidate,
and at least two light jets. If an event contains a reconstructed hadronic $\tau$-lepton candidate (\tauhadvis, as defined in \Sect{\ref{sec:objects}}), it is not included in the analysis, thereby removing the overlap with $HH$, $SH$ analyses with a $\tau\tau$ pair in the final state~\cite{HDBS-2019-27,HDBS-2022-44}.

Events with exactly one $b$-tagged jet and exactly two leptons with relaxed
requirements on the impact parameter $|d_0|/\sigma_{d_0}$, on the identification and isolation,
are used to define control regions to estimate the fake-lepton background.

The kinematic signature of the $HH \to b\bar{b}WW^* \to b\bar{b}\ell\nu q\bar{q}^\prime$ process differs significantly from the main $t \bar{t} \to b\bar{b}WW^* \to b\bar{b}\ell\nu q\bar{q}^\prime$ background.
In particular, $b$-jets and light jets pair to different particles in the two cases. The mass of the intermediate state particles is used to find the best match of each jet to the mother particle and
to test the compatibility of the observed event with the $HH$ or the $t \bar{t}$ process. This feature is exploited in two ways: using kinematic variables and using a kinematic fit. Two variables are constructed, called Higgsness ($H$) and
Topness ($T$), as originally proposed in Ref.~\cite{HiggsnessTopness} for the $H \to WW^* \to \ell \nu \ell \nu$ final state and modified for the $ H\to  WW^* \to \ell \nu q \bar{q}^\prime$ final state in Ref.~\cite{ATL-PHYS-PUB-2019-040}, furthermore two kinematic fits are built, one for the $t \bar{t}$ hypothesis and a second for the $HH$ hypothesis.
The definition of the $H$ and $T$ variables are the following:

\begin{scriptsize}
\[
\textbf{T}=\min\left(\lambda_{12}^{2},\lambda_{21}^{2}\right) \ \ \ \text{where:} \ \ \ \lambda_{ij}^{2}= \min\limits_{p_{z}^{\nu}}\left[\frac{\left(m_{b_{i}\ell\nu}^{2}-m_{t}^{2} \right)^2}{\sigma_{t}^{4}}+\frac{\left(m_{\ell\nu}^{2}-m_{W}^{2} \right)^2}{\sigma_{W}^{4}}+\frac{\left(m_{b_{j}qq}^{2}-m_{t}^{2} \right)^2}{\sigma_{t}^{4}}+\frac{\left(m_{qq}^{2}-m_{W}^{2} \right)^2}{\sigma_{W}^{4}}\right],
\]
\[
\textbf{H}= \min\limits_{p_{z}^{\nu}}\left[\frac{\left(m_{\ell\nu qq}^{2}-m_{H}^{2} \right)^2}{\sigma_{H}^{4}}+\min \left( \frac{\left(m_{\ell \nu}^{2}-m_{W}^{2} \right)^2}{\sigma_{W}^{4}}+\frac{\left(m_{qq}^{2}-m_{W_{peak}^{*}}^{2} \right)^2}{\sigma_{W^{*}}^{4}} , \frac{\left(m_{qq}^{2}-m_{W}^{2} \right)^2}{\sigma_{W}^{4}}+\frac{\left(m_{\ell\nu}^{2}-m_{W_{peak}^{*}}^{2} \right)^2}{\sigma_{W^{*}}^{4}}\right)\right],
\]
\end{scriptsize}

where $q$ indicates a light jet, $b$ a $b$-jet, $\ell$ the lepton, $\nu$ the neutrino, $m_t$ the mass of the top-quark, $m_W$ the mass of the $W$ boson and $m_{W_{peak}^*}$ is set to 40~\GeV, corresponding to the
value where the $m_{\ell \nu}$ distribution shows a maximum in the $HH$ case. The transverse components of the neutrino momentum are set to the MET value while its longitudinal component, $p_z^{\nu}$, is obtained by minimising the function. Its starting value before the minimisation process is obtained by imposing the $W$ mass constraint in the $t \bar{t}$ case and the Higgs boson mass constraint in the $HH$ case. In fact, in the $HH$ hypothesis, it is not possible to determine whether the on-shell $W$ decays to a quark/anti-quark  or to a lepton/neutrino pair. The $\sigma$ values used in the formulae are determined by fitting, with a Gaussian distribution,  the invariant mass reconstructed from the final state objects as predicted by the Monte Carlo simulation.
To reconstruct the $W$ boson which decays hadronically, the two \pT-leading light jets are used to obtain the invariant mass distribution. The pairing of the $b$-jet, the light jets and the lepton to the top mass for the $T$ variable, and of the light jets and the lepton to the on-shell and off-shell $W$ bosons is performed by minimising the $T$ and $H$ variables on all possible combinations.

\begin{figure}

\begin{center}
\subfloat[]{\includegraphics[width=0.33\textwidth]{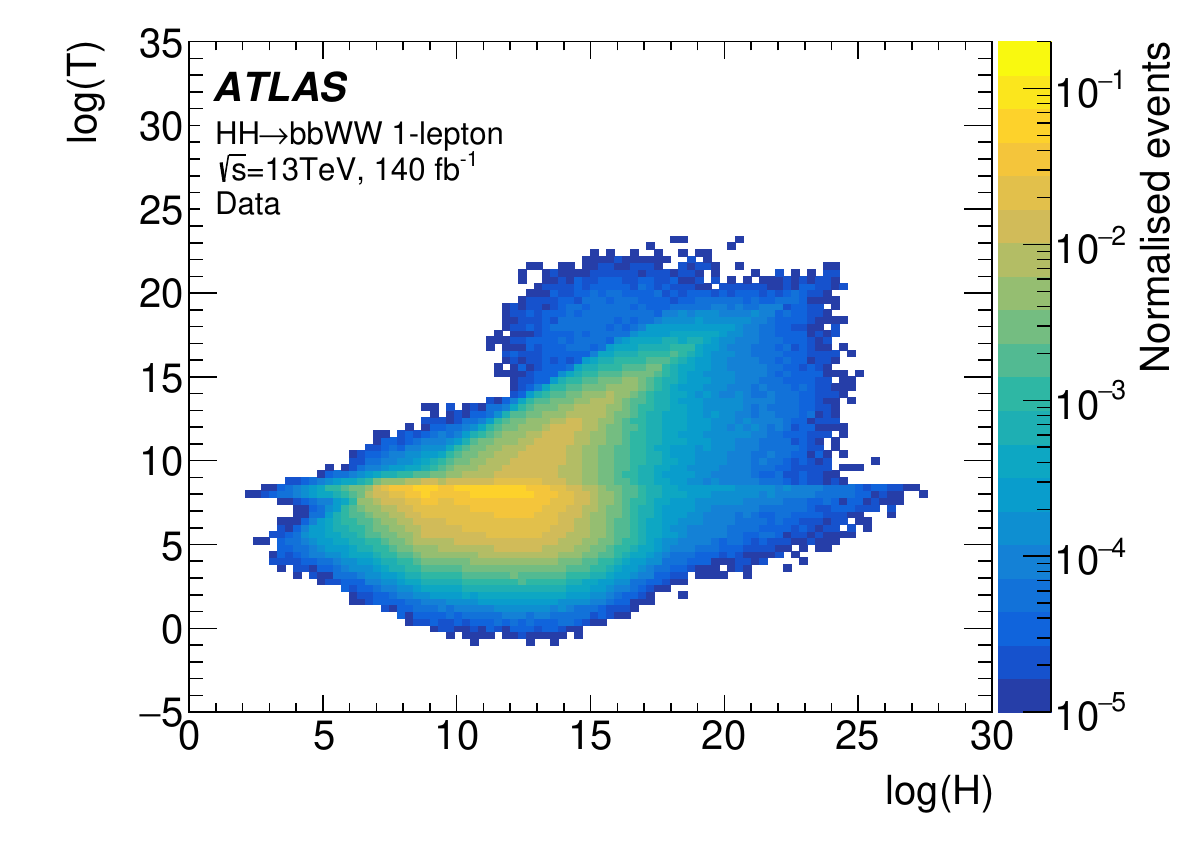}}
\subfloat[]{\includegraphics[width=0.33\textwidth]{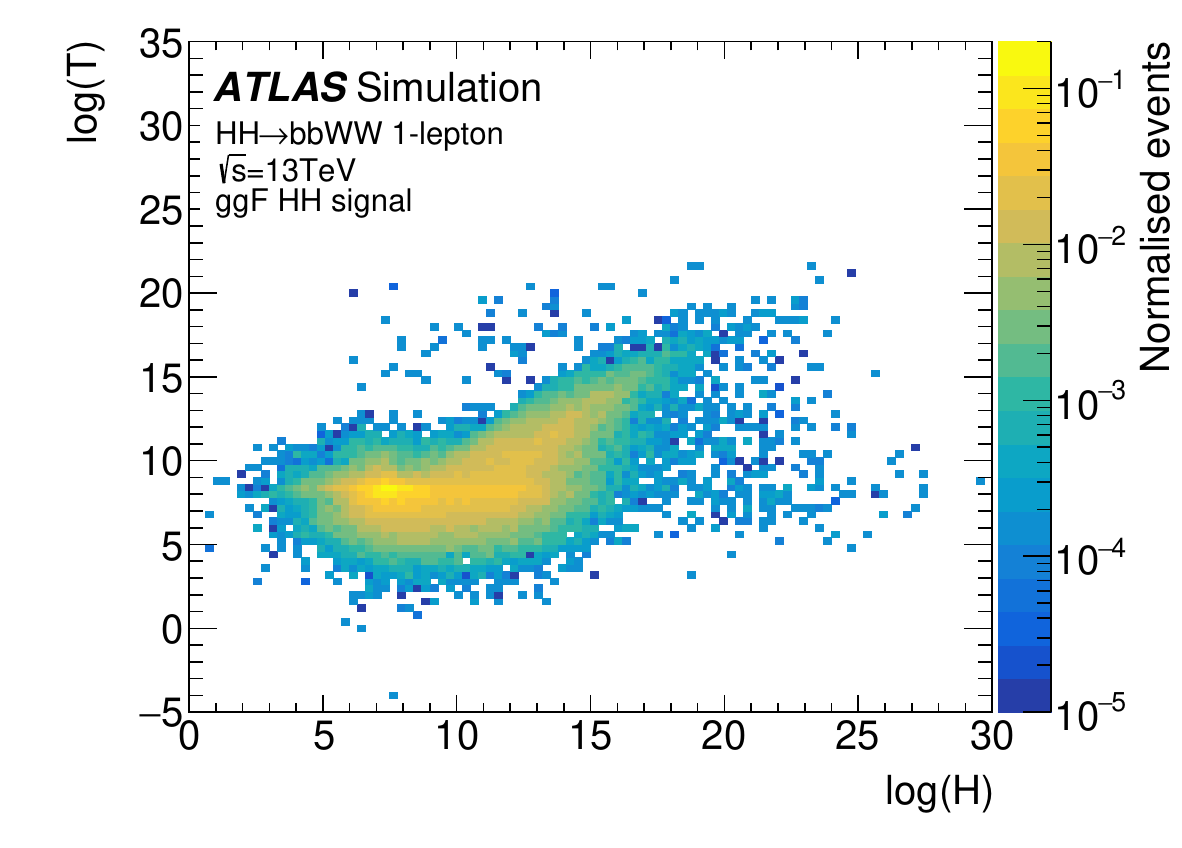}}
\subfloat[]{\includegraphics[width=0.33\textwidth]{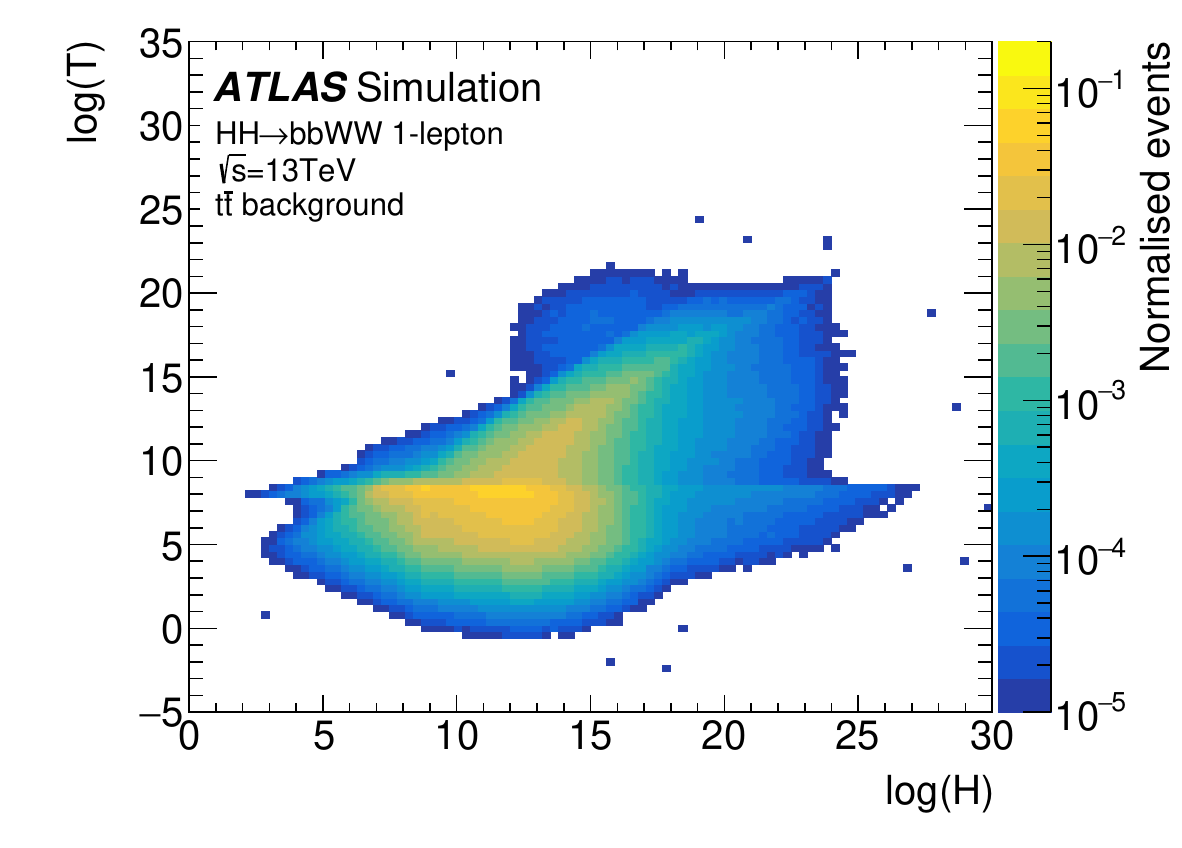}}
\\
\subfloat[]{\includegraphics[width=0.33\textwidth]{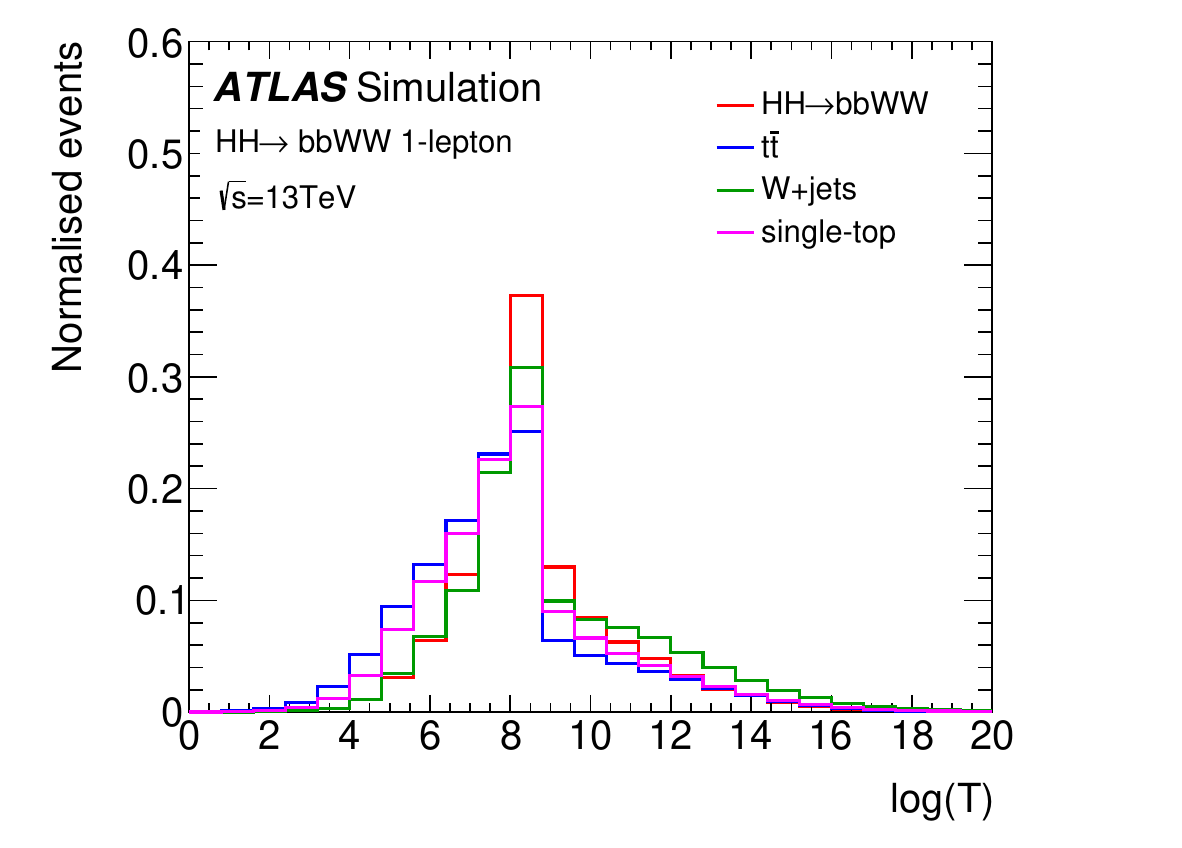}}
\subfloat[]{\includegraphics[width=0.33\textwidth]{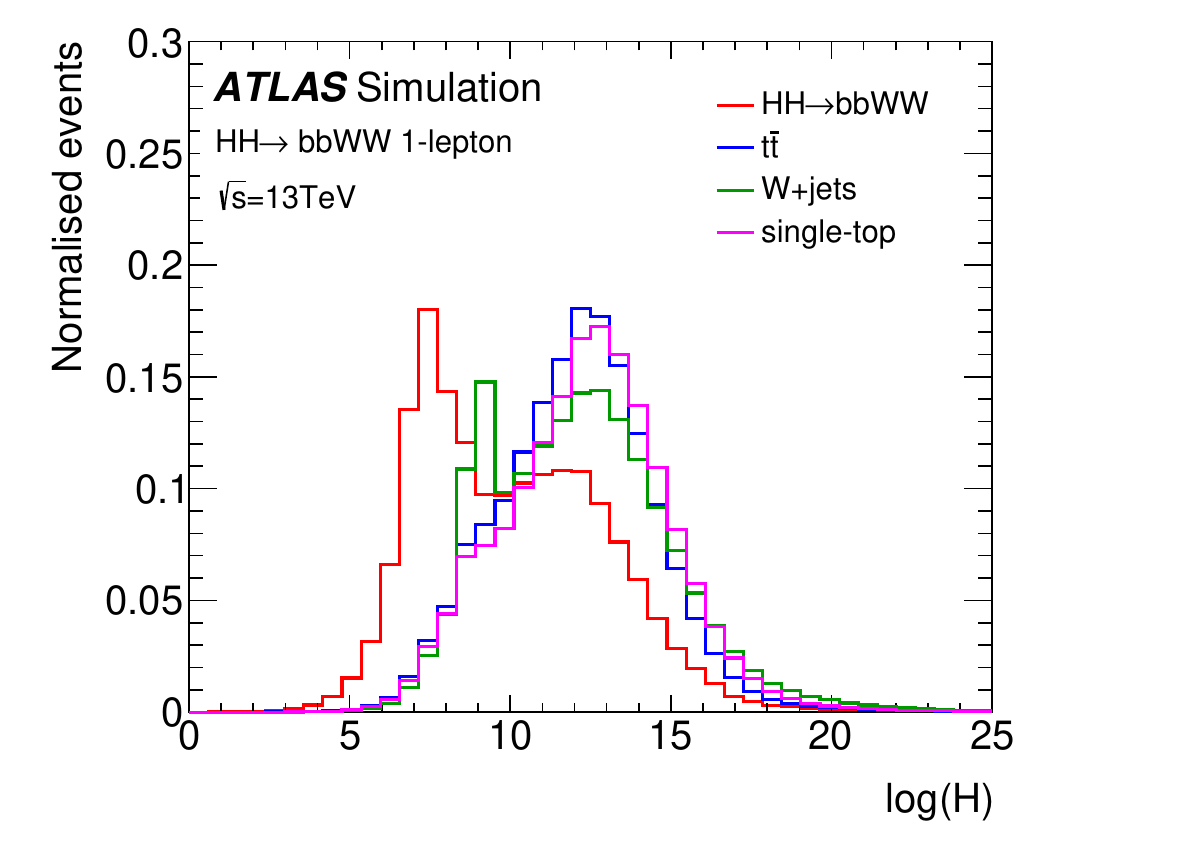}}
\end{center}
\caption{Distribution of the  Higgsness and Topness variables in a two-dimensional plane ($\log(H)$, $\log(T)$) for: (a) data, (b) $HH$ signal,  (c) $t\bar{t}$ background; and 1D projections of: (d) Topness variable and (e) Higgsness variable. Distributions are shown  for $HH$ signal, $t\bar{t}$ and $W+$jets backgrounds overlaid.}
\label{fig:THfigure}
\end{figure}

The $T$ and $H$ variables discriminate well  between the $t\bar{t}$ and $HH$ processes. Distributions of their natural logarithms [log$(T)$, log$(H)$]  are shown in \Fig{\ref{fig:THfigure}}. The accumulation of events around log$(T)=8$ is due to the kinematic properties of the events and the use of a fourth power in the definition of $T$.

The kinematic fit uses a likelihood function $(\mathcal{L})$~\cite{ERDMANN201418} that includes the energy response functions of jets, leptons and missing transverse energy and constraints on the mass of the top-quark and the $W$ boson in the $t \bar{t}$ hypothesis and on the Higgs boson in the $HH$ hypothesis. The minimum of the negative log-likelihood $[\Lambda = -\log(\mathcal{L})]$  is retained for each hypothesis as $\Lambda_{HH}$, $\Lambda_{t \bar{t}}$.

\begin{figure}
\begin{center}
\subfloat[]{\includegraphics[width=0.45\textwidth]{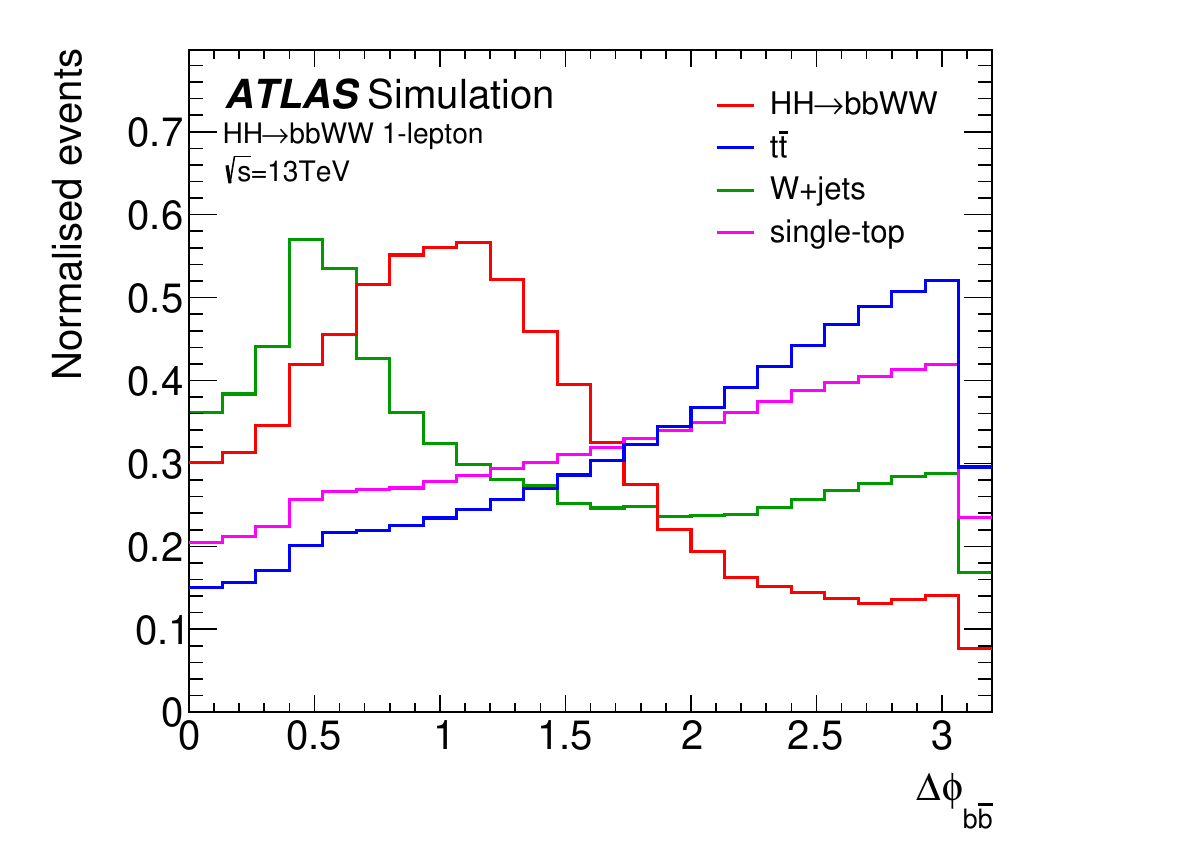}}
\subfloat[]{\includegraphics[width=0.45\textwidth]{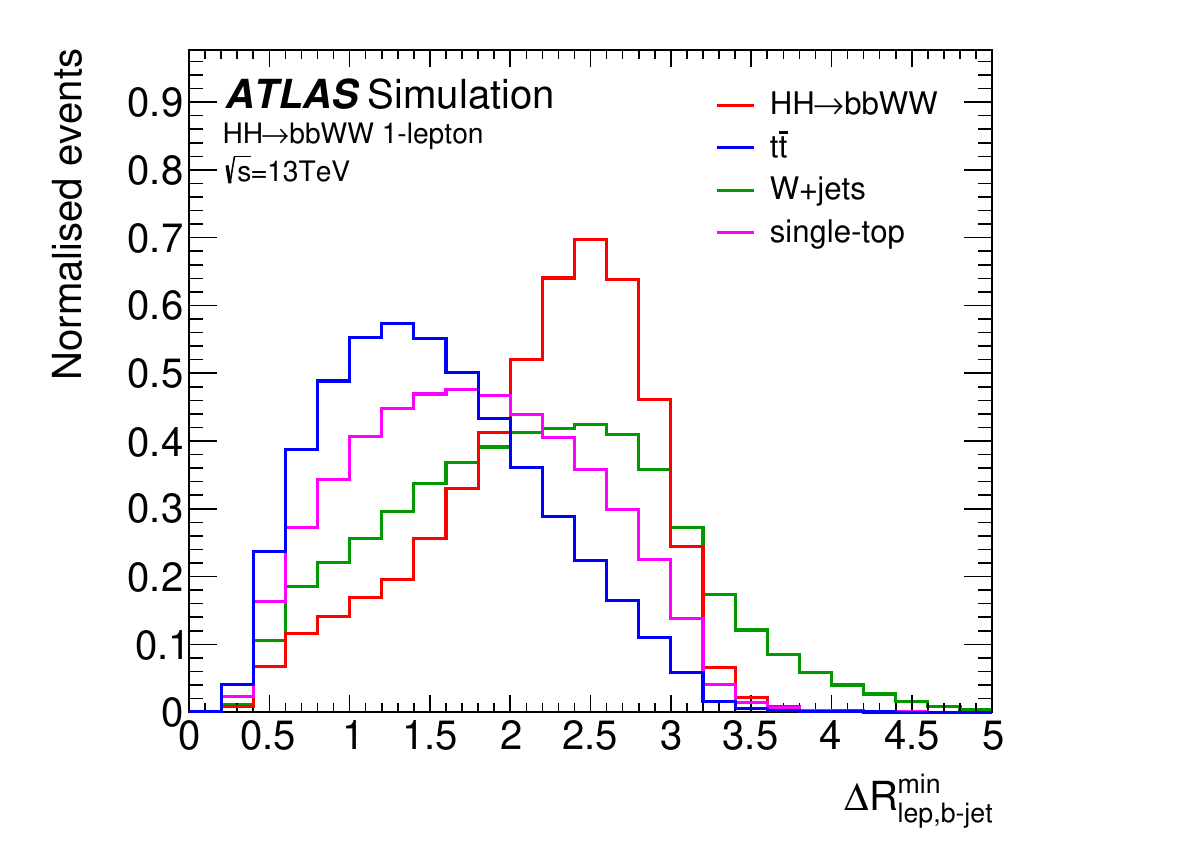}} \\
\subfloat[]{\includegraphics[width=0.45\textwidth]{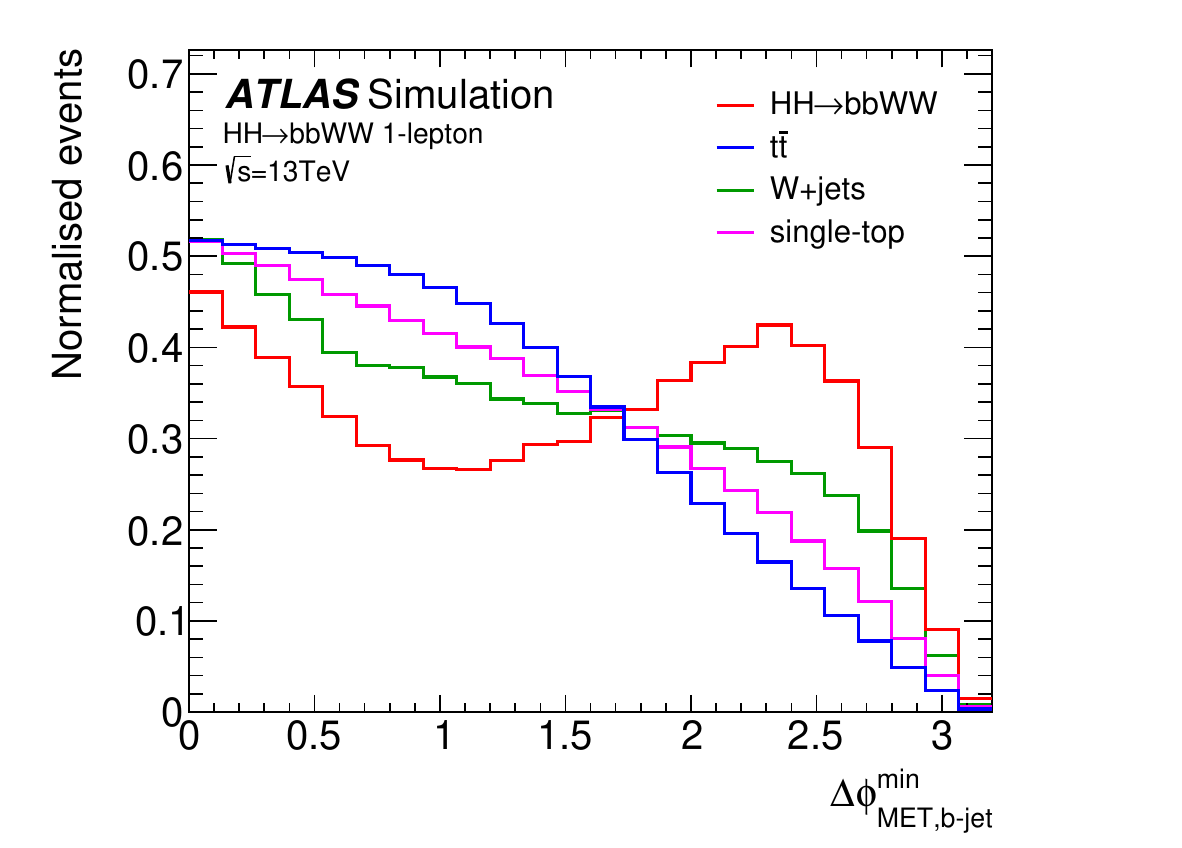}}
\subfloat[]{\includegraphics[width=0.45\textwidth]{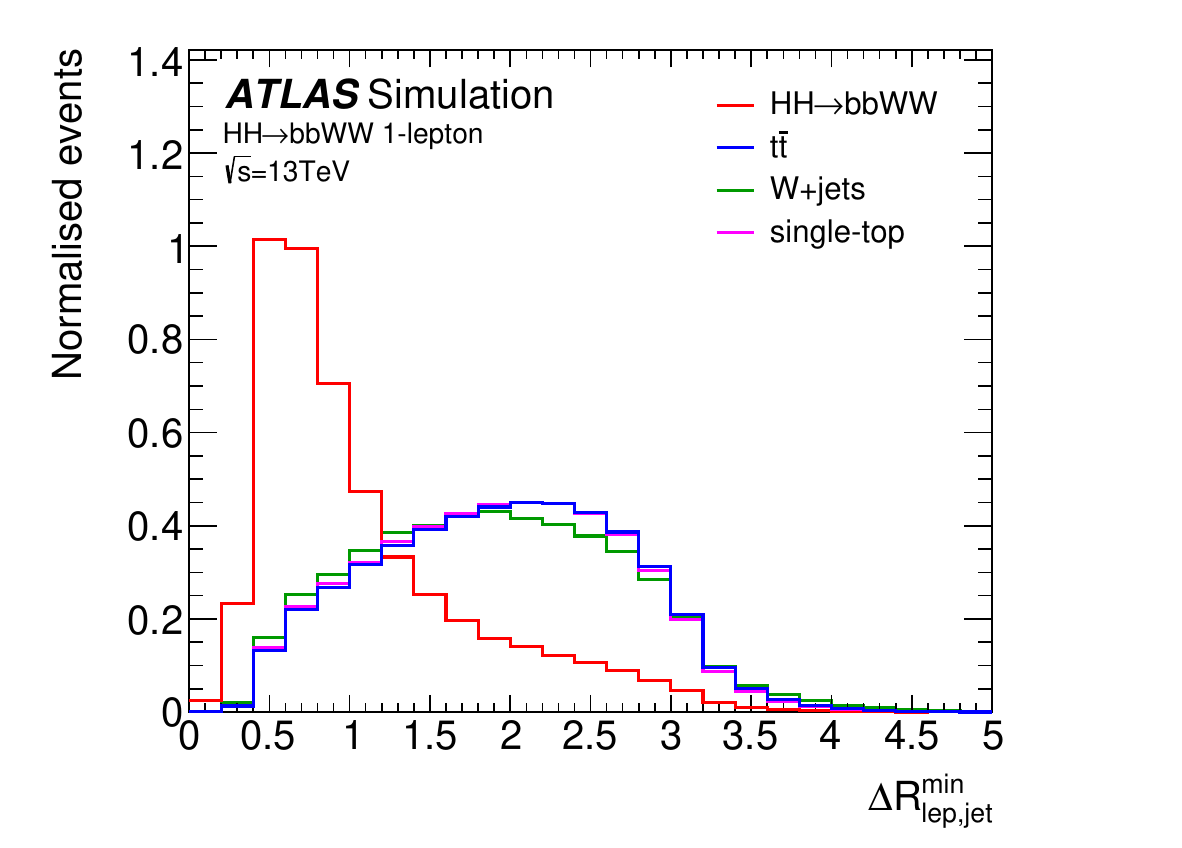}}
\end{center}
\caption{Distribution of the most discriminant angular variables characterising the jets, the lepton and the MET in the event: (a) azimuthal angular distance of the two $b$-jets, $\Delta \phi_{b \bar{b}}$;  (b) minimum $\Delta R$  between the lepton and the $b$-jets, $\Delta R_{\textrm{lep},b\textrm{-jet}}^{\textrm{min}}$; (c) the minimum azimuthal angular distance  between the MET and the $b$-jets,  $\Delta\phi_{\textrm{MET},b\textrm{-jet}}^{\textrm{min}}$; (d) minimum $\Delta R$ between the lepton and the light jet,  $\Delta R_{\textrm{lep,jet}}^{\textrm{min}}$.} \label{fig:variables}
\end{figure}

\begin{figure}
\begin{center}
\subfloat[]{\includegraphics[width=0.45\textwidth]{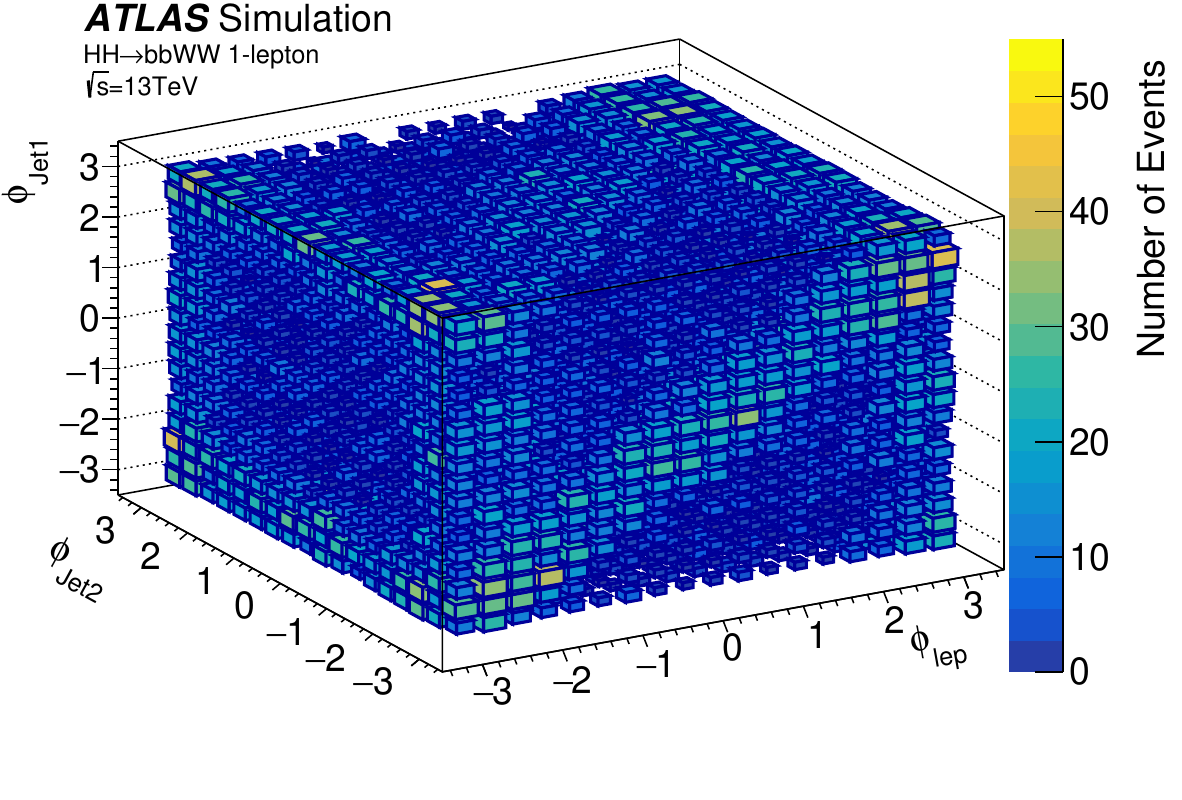}}
\hspace{3mm}
\subfloat[]{\includegraphics[width=0.45\textwidth]{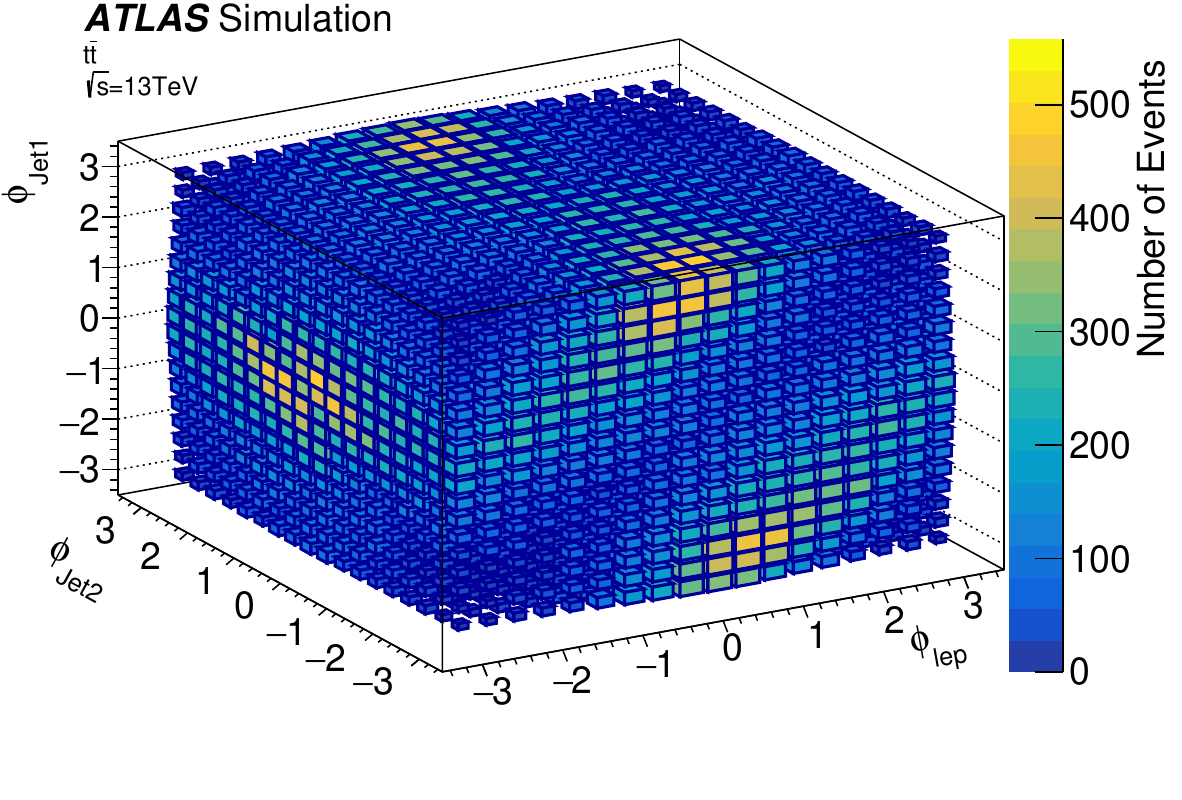}} \\
\subfloat[]{\includegraphics[width=0.45\textwidth]{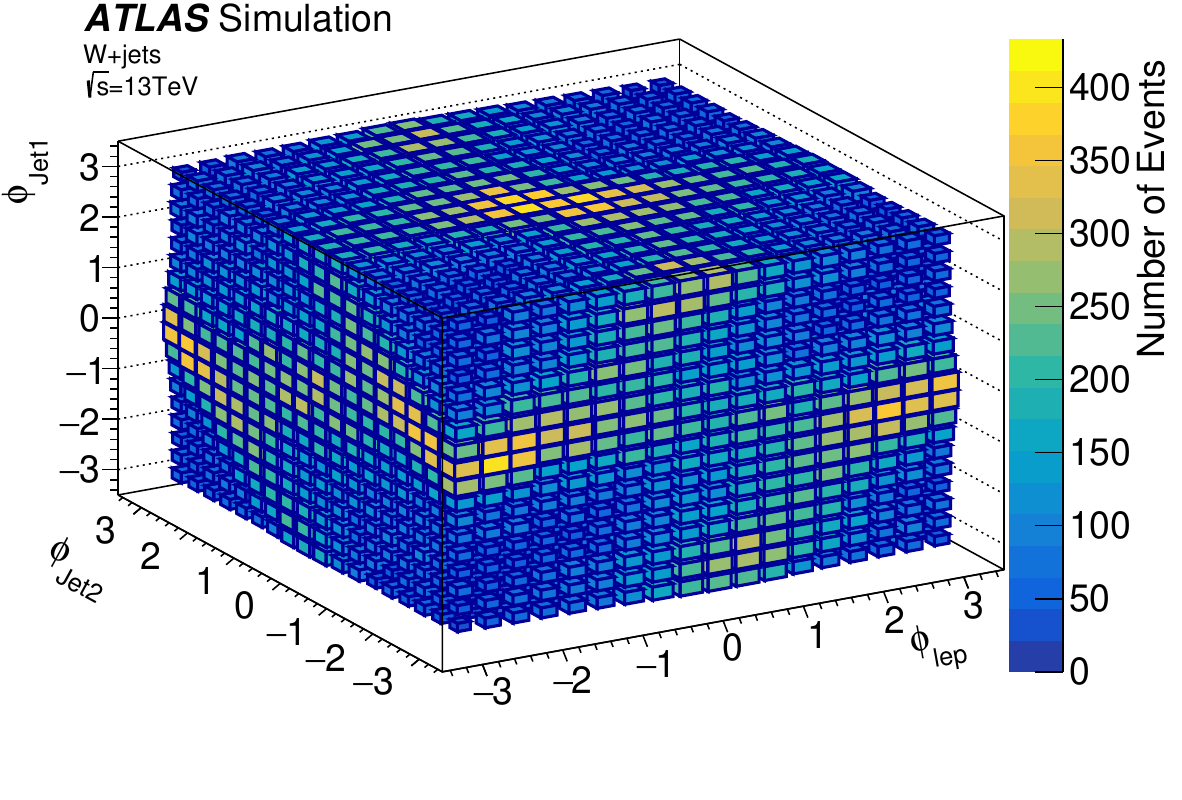}}
\caption{The three-dimensional distribution of the azimuthal angle $\phi$ of the two light jets with highest \pt, $\phi_{\mathrm{Jet1}}$ and $\phi_{\mathrm{Jet2}}$ and of the lepton $\phi_{\mathrm{lep}}$ for the: (a) signal, (b) $t\bar{t}$ and (c) $W+$jets processes.} \label{fig:variable_correlation}
\end{center}
\end{figure}

In addition to the intermediate mass values, the $HH$ and $t \bar{t}$ backgrounds differ for many other kinematic distributions. To profit from the two different decay topologies, many variables are exploited and used in the Neural Network algorithm described in Sec.~\ref{sec:NN}:

\begin{enumerate}
\item log$(T)$, log$(H)$, $\Lambda_{HH}$, $\Lambda_{t \bar{t}}$;
\item the jet multiplicity of the event, i.e. the total number of light and $b$-jets;
\item the flavour of the signal lepton;
\item $p_{\text{T}}$, $\eta$ and $\phi$ of the two $b$-jets, the three light jets with the highest $p_{\text{T}}$ and the lepton;
\item the missing transverse energy magnitude $E_{\text{T}}^{\textrm{miss}}$, its $\phi$ angle and the angular distance between the lepton and the \met direction $\Delta\phi_{\ell,\mathrm{MET}}$;
\item the invariant mass, $m_{b \bar{b}}$, and the angular distances $\Delta R_{b\bar{b}}$ and $\Delta \phi_{b \bar{b}}$ of the two $b$-jets;
\item the transverse mass of the lepton-MET pair, $m_{\text{T}}^{\ell \nu}$, the $WW$ system, $m_{\mathrm{T}}^{WW}$, and the $HH$ system, $m_{\mathrm{T}}^{HH}$ defined as :
\[
m_\text{T} = \sqrt{\left( \sum_i E_\text{T}^i \right )^2 - \left | \sum_i \vec{p}_\text{T}^{i} \right |^2}
\]
where $\vec{p}_T$ is the transverse momentum of the particle and $E_{\text{T}} = \sqrt{|\vec{p}_{\text{T}}|^2+m^2}$.  The sum is computed over the particles belonging to the system: for $m_{\text{T}}^{\ell \nu}$  the system includes the lepton and the MET, for  $m_{\mathrm{T}}^{WW}$ the $W$ decaying hadronically ($W_{\mathrm{had}}$) is added to the system. Finally, for  $m_{\mathrm{T}}^{HH}$ the $b$-jets are also added to the system. The  $W_{\mathrm{had}}$ momentum is obtained from the two light-jets closest in $\Delta R$ among the three leading ones;

\item the minimum and maximum $\Delta R$ between the lepton and the jets ($\Delta R_{\textrm{lep,jet}}^{\textrm{max}}$, $\Delta R_{\textrm{lep,jet}}^{\textrm{min}}$,  $\Delta R_{\textrm{lep},b\textrm{-jet}}^{\textrm{max}}$, $\Delta R_{\textrm{lep},b\textrm{-jet}}^{\textrm{min}}$) and the minimum and maximum $\Delta\phi$ between the MET  and the two $b$-jets  ($\Delta\phi_{\textrm{MET,}b\textrm{-jet}}^{\textrm{min}}$, $\Delta\phi_{\textrm{MET},b\textrm{-jet}}^{\textrm{max}}$).
These variables exploit the different angular distributions of leptons and jets between  the $HH$ signal and the two main backgrounds, $W$+jets and $t \bar{t}$.
In fact, in the $HH$ signal, for high values of $m_{HH}$, the lepton and the light jets are quite close to each other, being  generated by the same Higgs boson, contrary to the $t\bar{t}$ and $W$+jets backgrounds, where light jets are produced mainly through initial and final state radiation. Furthermore, the neutrino momentum and the $b$-jet momenta are close in $\Delta \phi$ in the $t\bar{t}$ background case and are far apart in the $HH$ case.
\end{enumerate}
\Fig{\ref{fig:variables}} shows the most discriminating variables between the $HH$ signal and the $t\bar{t}$, $W$+jets and single-top backgrounds.
One particularly interesting feature is the different correlation pattern among the basic variables characterising the jets and the lepton. An example is shown in \Fig{\ref{fig:variable_correlation}}, where the
azimuthal angle of the two jets with highest \pt and of the lepton is shown. In the signal case the events populate the region of
the phase space with similar $\phi$ values. The use of Neural Networks allows to exploit the  different correlation patterns of the observables to increase the separation between signal and background, as described in the next section.

\subsection{Neural Network based discriminants}
\label{sec:NN}

To exploit the potential of the several discriminating variables and their correlations, several  neural networks (one for $HH$ and several for $SH$ to address the wide range of signal model parameters considered, see \Sect{\ref{sec:SHanalysis}} for details) are built with a  multi-output classifier  trained against the $HH$ and the $SH$ signals  and the three  main background contributions: $t\bar{t}$, $W$+jets and single-top.
For single-top,  only the $Wt$ channel is used in training because the $s-$channel and the $t-$channel processes show  similar distributions to the $t\bar{t}$ background.

The discriminant architecture is common for all cases, using a deep neural network model (DNN)  built using the Keras library~\cite{chollet2015keras} and TensorFlow~\cite{abadi2016tensorflow} as a backend.
The network inputs are first standardised with the use of Scikit-Learn tools~\cite{scikit-learn} to transform the variables to standard normally distributed data.
The input standardised variables are then passed through five fully connected layers, also referred to as dense layers.
The first four layers have 512 nodes, while the last one has 256 nodes.
Between each pair of dense layers there is a dropout layer~\cite{Srivastava:2014}, and an additional dropout layer is present between the last hidden layer and the output layer.
The architecture of the model is completed with a final layer with four output nodes activated by the softmax activation function.

The activation functions used for the dense layers are rectified linear units (ReLU).
The softmax activation function of the output layer allows for the interpretation of the outputs as
the probability that the event is one of the four processes, as the sum over all outputs is equal to unity.
In the following, the outputs of the DNN are denoted as $p_{HH}$ ($p_{SH}$), $p_{t\bar{t}}$, $p_{W}$, and $p_{Wt}$,  for each of the classified processes.
A total of 40 input variables (described in the previous section) is provided  to the DNN.

The samples used in the training are split into  independent training, validation for hyperparameter optimisation and a test set.
The cross-validation method with five folds is adopted in the DNN training to avoid any bias in the response of the DNN in the analysis:
the initial set of events is split into five random subsamples (``folds''), and the
training is performed five times, using each time a different fold as the training sample and the remaining folds for the training and the validation datasets.

The output probabilities are combined to obtain a better discriminating variable, $d_{HH}$ (or $d_{SH}$). This variable is defined as the logarithm of the ratio between the probability of the event to be a $HH/SH$ signal and the probability of being one of the three main backgrounds, $t\bar{t}$, $W$+jets and single-top:

\begin{equation}
d_{HH/SH} = \log\left( \frac{p_{HH/SH}}{p_{t\bar{t}}+p_{W}+p_{Wt}} \right) = \log \left( \frac{p_{HH/SH}}{1 - p_{HH/SH}} \right)
\label{eq:dHH}
\end{equation}

\begin{figure}[!h]
\begin{center}
\subfloat[]{
\includegraphics[width=.373\textwidth]{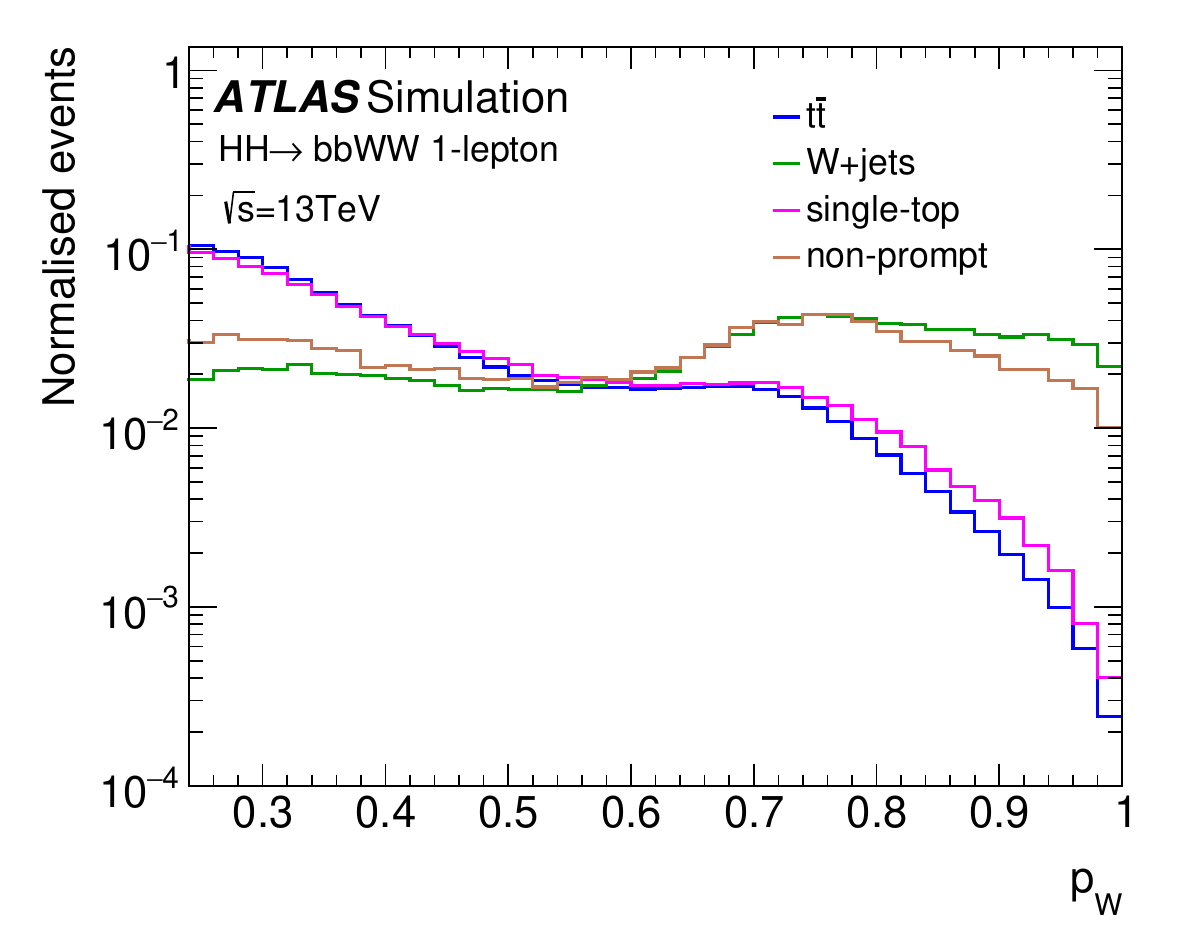}
\label{fig:DNN_pW}
}
\subfloat[]{
\includegraphics[width=.41\textwidth]{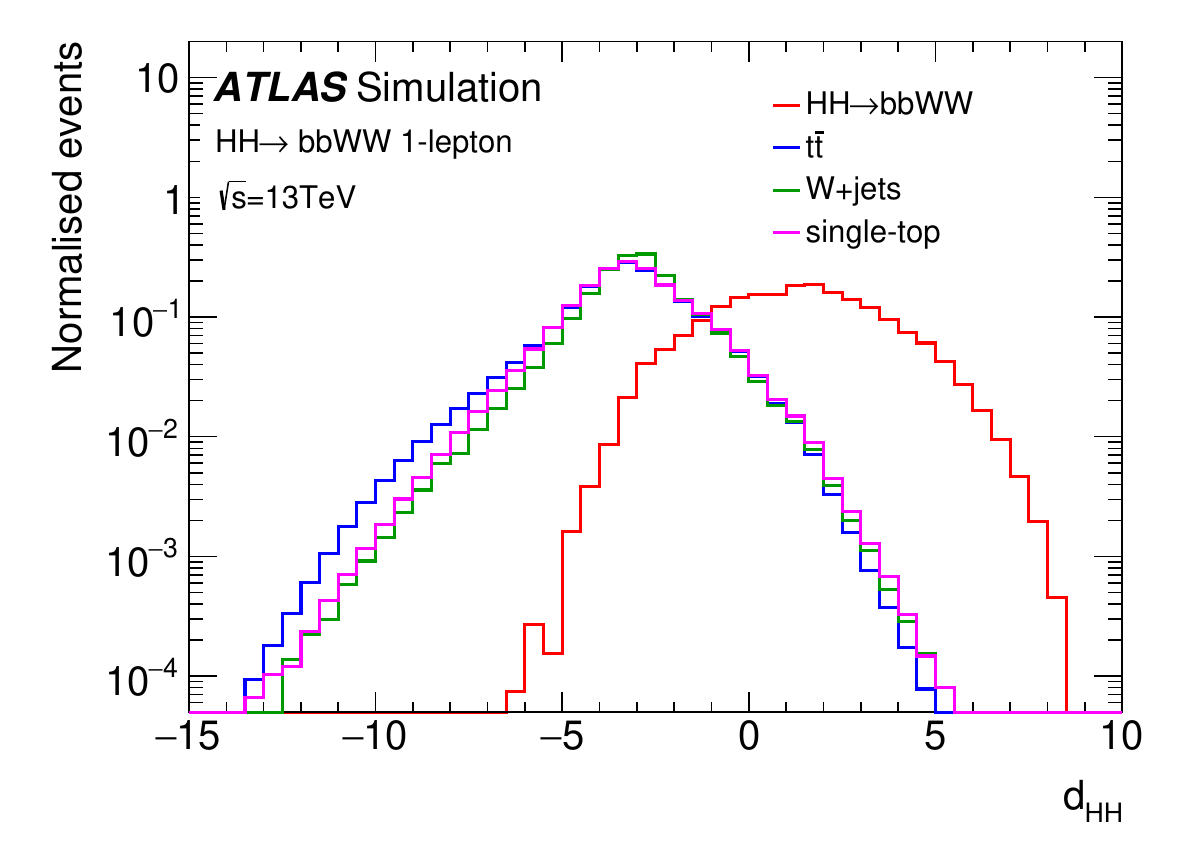}
\label{fig:DNN_dHH_fold1}
}
\end{center}
\caption{Shape of (a) $p_{W}$ and (b) $d_{HH}$ discriminating variable for signal and background events.
Entries are normalised to have unit area for each sample distribution.}
\label{fig:DNN_dHH}
\end{figure}

\begin{figure}
\centering
\includegraphics[width=1\textwidth]{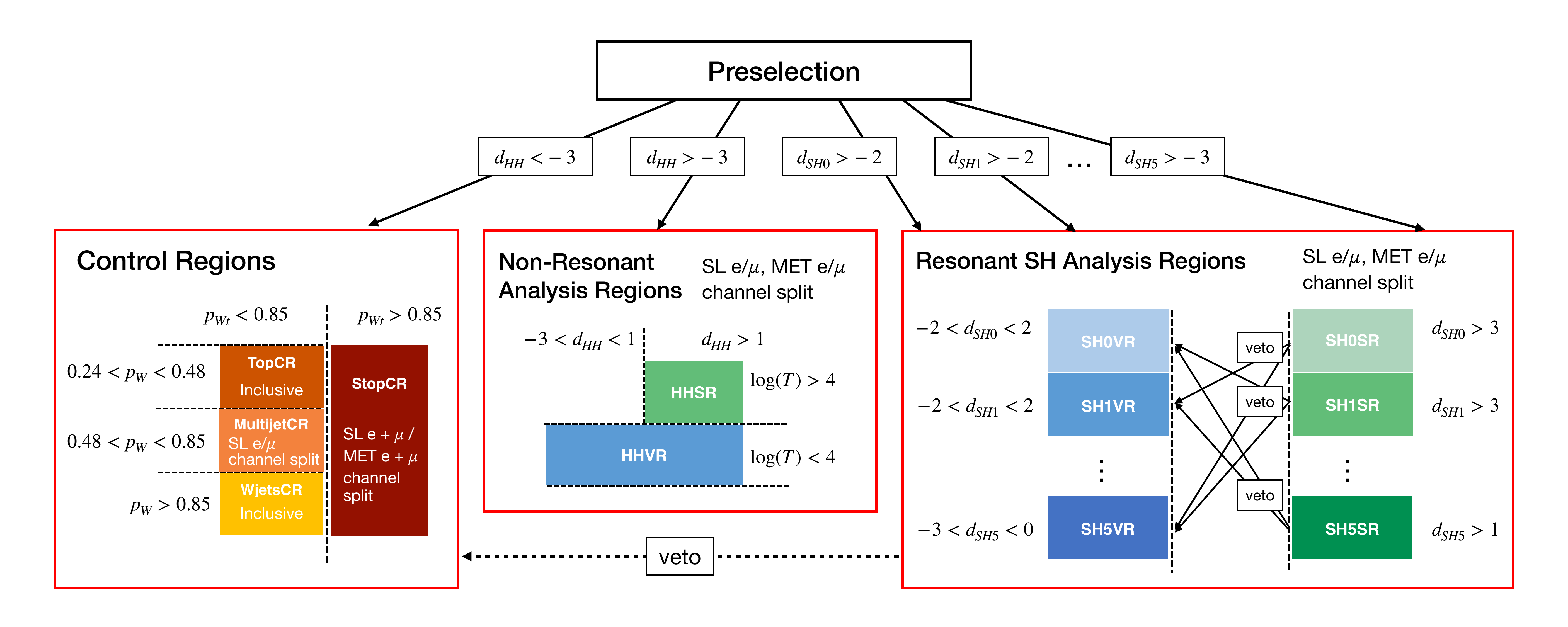}
\caption{Visual representation of the analysis categorisation and region definitions for the $HH$ non-resonant analysis and for the $SH$ analysis. TopCR and WjetsCR are kept inclusive, StopCR is split into single-lepton and MET trigger channels, and MultijetCR is split into electron and muon channels. Non-resonant VRs and SRs (denoted by HHVR and HHSR) and $SH$ VRs and SRs (denoted by SH0VR-SH5VR and SH0SR-SH5SR) are split based on the trigger type and the lepton flavour. The cuts on the Neural Network (NN) discriminating variables $d_{HH}$ (for the non-resonant analysis), $d_{SH0}$, $d_{SH1}$, $d_{SH2}$, $d_{SH3}$, $d_{SH4}$ and $d_{SH5}$ (for the $SH$ analysis) are not orthogonal by definition, hence explicit vetoes need to be applied to SH CRs and VRs (see Sec. 5.3).}
\label{fig:regions}
\end{figure}

The distributions of $p_{W}$ and $d_{HH}$ for the $HH$ analysis are shown in \figurename~\ref{fig:DNN_dHH}.
The DNN output probabilities and the $d_{HH}$ variable are used to define mutually exclusive signal regions (SRs), validation regions (VRs),
and control regions (CRs), summarised in Figure~\ref{fig:regions}.
The SRs and VRs are further split depending on the type of trigger that selected the events, single-lepton (SL) or MET triggers,
and on the flavour of the lepton: electron ($e$) or muon ($\mu$).
CRs are built to define regions with enhanced contribution of the major backgrounds: TopCR for $t\bar{t}$, WjetsCR for $W$+jets, StopCR for single-top, and MultijetCR for multijet background.
TopCR and WjetsCR are kept inclusive, while StopCR and MultijetCR are split into two regions, respectively depending on the trigger type (SL/MET) and on the lepton flavour ($e/\mu$). The splitting is performed in this way to improve the sensitivity of the analysis, with the electron and muon channels being affected by different fake background contributions, and with MET-triggered events characterised by
different background composition than the lepton-triggered events.

\subsection{Event classification for the $SH$ analysis}
\label{sec:SHanalysis}
\begin{figure}[htbp]
\centering
\includegraphics[width=1\textwidth]{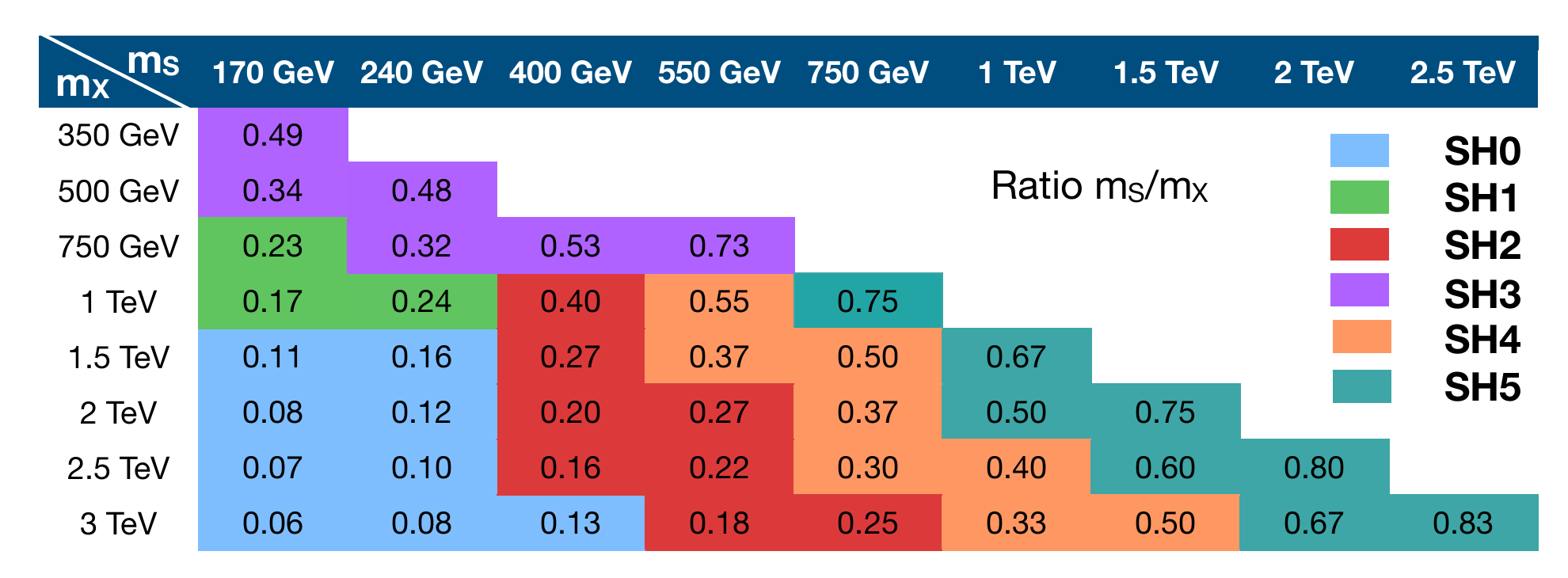}
\caption{Schematic representation of the $SH$ mass point groups. The same colour is used for the mass points belonging into the same kinematic group. The mass ratio $m_S/m_X$ of each mass point is shown.}
\label{fig:shgroups}
\end{figure}

The $SH$ analysis probes 42 mass points in a $(m_S,m_X)$ grid, using the simulated samples with mass values tabulated in
\Fig{\ref{fig:shgroups}}.
The considered signals are arranged into six groups (SH0-5) shown in \figurename~\ref{fig:shgroups}. Within each group, the signals share similar kinematic properties, based mainly on the boost of the $S$ and $H$ decay products. A dedicated DNN discriminator is trained for each group to maximise the analysis sensitivity while reducing the number of trained networks significantly.
Six sets of SRs are designed to target different $SH$ mass groups, selecting different intervals of the DNN discriminating variable $d_{SHi}$ ($i=0-5$). Each set contains four SRs according to the trigger (MET or lepton) and to the lepton type (electron or muon).
Each $SH$-SR set has corresponding CRs defined starting from the CRs of the non-resonant $HH$ analysis and removing events that are part of the $SH$-SR set.
Thresholds on the DNN discriminating variables $d_{SHi}$  are chosen to keep background contamination under control with minimum
loss of sensitivity. Validation regions are defined applying sideband cuts on the DNN discriminating variables and vetoing events satisfying all SR selections.

\begin{figure}[htbp]
\centering
\subfloat[]{
\includegraphics[width=0.45\textwidth]{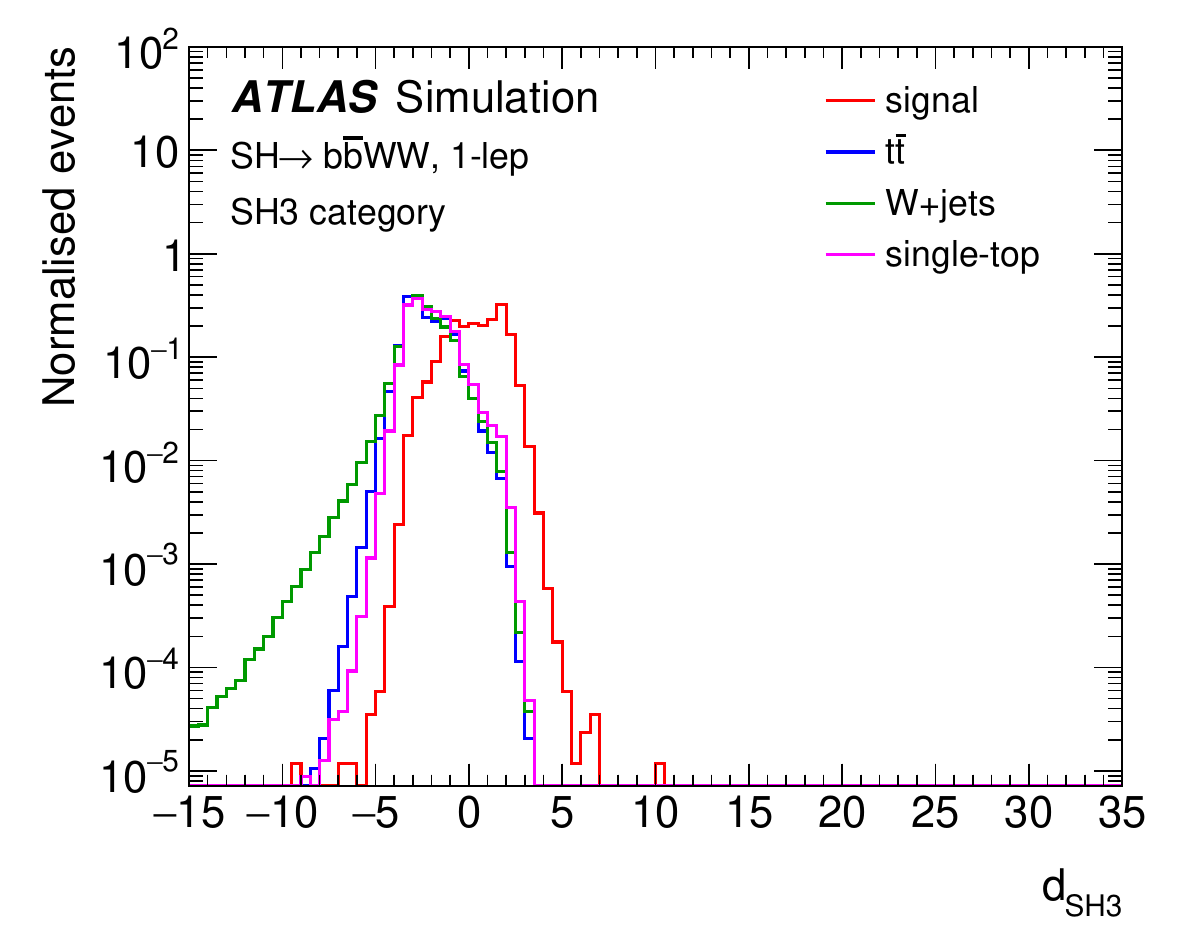}
\label{fig:sh_sr3dsh}
}
\subfloat[]{
\includegraphics[width=0.45\textwidth]{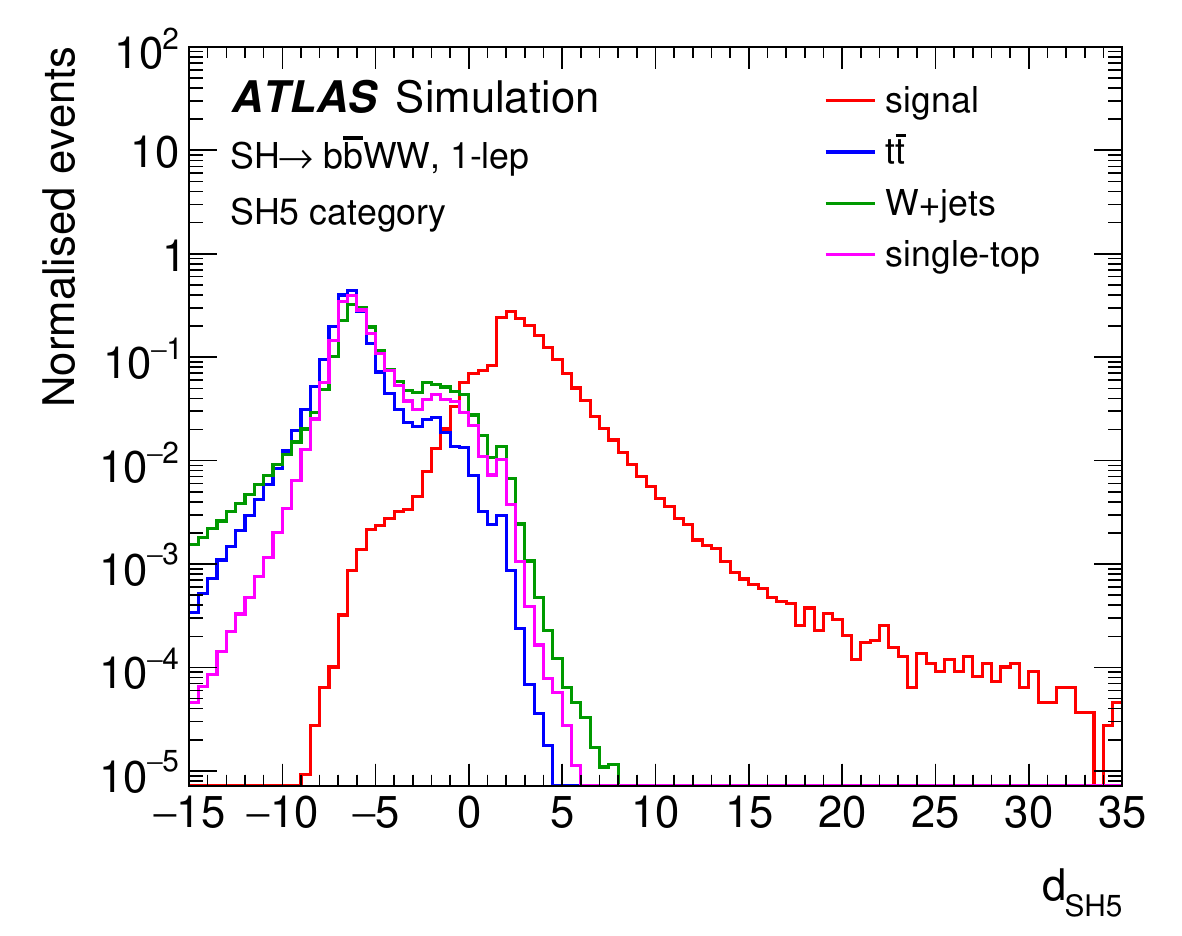}
\label{fig:sh_sr5dsh}
}
\caption{Distribution of (a) $d_{SH3}$ and (b) $d_{SH5}$ discriminating variable in corresponding SR. The $SH$ signal denotes a mixture of mass points being targeted by a particular category.
}\label{fig:sh35_dsh}
\end{figure}

The SR, CR and VR definitions for the $SH$ analysis are shown in Figure~\ref{fig:regions}. The ranges of the SR and VR discriminating variables are summarised in Table~\ref{tab:sh_srvr_binning}. The distributions of the discriminating variables are shown in \Fig{\ref{fig:sh35_dsh}} for two representative regions: SH3 and SH5.
The resolved selection investigated here is expected to be particularly sensitive to models with $m_X < 1$~\TeV. For $m_X > 1$~\TeV, the two $b$-jets are difficult to resolve and better reconstructed as a single Large-$R$ jet.
To allow for a future statistical combination with a dedicated boosted analysis, events are removed if they pass the following criteria: the Large-$R$ jet multiplicity is 1 or 2, one of the Large-$R$ jets has $\Delta R<1.0$ to a small-$R$ $b$-jet, the leading Large-$R$ jet \pt is greater than 450~\GeV. The impact of these selections on the resolved analysis is found to be negligible in the relevant range $m_X < 1$~\TeV.

\begin{table}[!h]
\caption{$HH$ and $SH$ region definitions and binning of the discriminant variables.}
\label{tab:sh_srvr_binning}
\begin{center}
\begin{tabular}{l|c||c|l|l}
\toprule \toprule
NN & VR definition &  SR definition &  SR binning SL & SR binning MET  \\
\midrule
HH & $d_{HH} > -3$ and $\log(T)<4$ & $d_{HH} > 1$ and $\log(T)>4$ & [1,2,3,4,5,6,$\infty$) & [1,2,3,4,5,$\infty$)   \\

SH0 & $-2 < d_{SH0} < 2$ & $d_{SH0} \geq
3$ & [3,4,5,$\infty$) & [3,$\infty$)   \\
SH1 & $-2 < d_{SH1} < 2$ & $d_{SH1} \geq
3$ & [3,4,5,6,7,$\infty$) & [3,4,5,$\infty$)   \\
SH2 & $-4 < d_{SH2} < 2$ & $d_{SH2} \geq
3$ & [3,4,$\infty$) & [3,$\infty$)   \\
SH3 & $-7 < d_{SH3} < -4$ & $d_{SH3} \geq
-1$ & [-1,0,1,2,3,$\infty$) & [-1,0,1,2,$\infty$)   \\
SH4 & $-4 < d_{SH4} < -2$ & $d_{SH4} \geq
-1$ & [-1,0,1,2,3,4,$\infty$) & [-1,0,1,2,$\infty$)   \\
SH5 & $-3 < d_{SH5} < 0$ & $d_{SH5} \geq
1$ & [1,2,3,4,$\infty$) & [1,2,$\infty$)  \\
\bottomrule \bottomrule
\end{tabular}
\end{center}
\end{table}


\section{Background estimation}
\label{sec:resolved_backgrounds}
Most background processes have a lepton in the final state, they are named prompt lepton backgrounds.
Additionally, a portion of the background originates from non-prompt leptons from the decay of heavy-flavour hadrons.
These contributions, collectively referred to as fake backgrounds, are less
accurately modelled by the MC simulation; therefore they are estimated with data-driven techniques.

The dominant background is the $t\bar{t}$ process; in addition there are smaller contributions
from $W$+jets and single-top.
The prompt lepton components of these backgrounds are modelled using simulation.
Dedicated control regions, as defined in \Sect{\ref{sec:selection}},  are used to normalise and validate the estimate.
Less relevant backgrounds, such as $Z$+jets and diboson are normalised to the expected SM cross-sections.

Fake backgrounds are estimated with the fake-factor method~\cite{EGAM-2019-01}.
The fake-factor is defined as $F = N_{\text{P}}/N_{\text{F}}$, where $N_{\text{P}}$
is the number of fake leptons that satisfy the object selection criteria outlined in \Sect{\ref{sec:objects}}.
$N_{\text{F}}$ is the number of fake leptons that do not fulfil all the selection criteria, but still satisfy a looser  baseline selection.
No
isolation requirement is applied for baseline electrons and muons and
the ``Loose'' (``Medium'')  electron (muon)  identification criterion is applied.
Furthermore, the electron crack region ($1.37 < |\eta| < 1.52$) is retained.
The selections on the impact parameters of electrons and muons are relaxed
to $|d_0|/\sigma(d_0)<10$ and $|\Delta z_0 \sin\theta|<50$\,mm.
Leptons matching a pile-up vertex inside  $0.5$\,mm range are excluded.
The fake factors $F_i$ are evaluated in bins of $\eta$ (barrel and endcap regions) and $p_\mathrm{T}$ in a fake-enriched control regions defined in the following.
The fake contribution in each signal region is estimated from data using the expression:
\begin{equation}
N_{\text{P}}^{\text{fakes}} =
\left[ \sum\limits_{i}F_i \times \left(N_{\mathrm{CR}i} - N^{\mathrm{prompt}}_{\mathrm{CR}i}\right) \right],
\label{eq:fake_factor_sum}
\end{equation}

The index $i$ runs over the bins used in the fake factor definition and CRi  are the respective regions defined using the SR selection criteria with the exception that loose leptons are used instead of signal region leptons. Simulated events are used to subtract the contamination of prompt leptons from the CR, referred to as $N^{\mathrm{prompt}}_{\mathrm{CR}i}$.

The fake-factor of a trigger-matched lepton differs significantly from
that of a  lepton that is not required to match the trigger condition.
This difference arises from the isolation requirement imposed by the single-lepton trigger, which removes a substantial fraction of loose-lepton events.
Dilepton events and single-lepton trigger are used to define fakes-enriched
regions from which two separate sets of fake-factors are derived: one
for trigger-matched leptons, used to compute the fakes for lepton
triggered events,  and another for  leptons not required
to match the trigger, used for  MET triggered events. Background from prompt leptons is subtracted from these regions before computing the fake factor. The prompt lepton contamination is particularly relevant in the tight region, where it reaches 70\% for the electron case. The uncertainty arising from the prompt leptons contamination subtraction (derived from MC normalised to data for the top background) induces a 100\% uncertainty on the fake factor.

One of the main sources of misidentification originates from semileptonic $b$-hadron decays.
The misidentification probability  decreases when jets containing $b$-hadrons are correctly reconstructed as $b$-jets,
consequently, the fake-factor is affected by the $b$-tagging requirement.
Therefore the fakes-enriched regions requires one $b$-jet, ensuring a high number of fakes while
maintaining a background composition similar to that of the signal region.
To suppress the contribution from prompt $Z$+jets events, same-flavour opposite-sign events are vetoed.
Additionally, a minimum of three light jets is required to match the jet multiplicity of the signal region.
A dedicated control region is defined to correct the prompt $t\bar{t}$ normalisation in the fakes-enriched control region.
This region is selected using a dilepton final state with two $b$-jets. The reconstructed leptons are
required to have opposite sign and different flavours (one electron and one muon). A normalisation factor of $0.834 \pm 0.004$ is obtained from a $t\bar{t}$ fit to data.

\begin{figure}[!h]
\begin{center}
\subfloat[]{\includegraphics[width=0.45\textwidth]{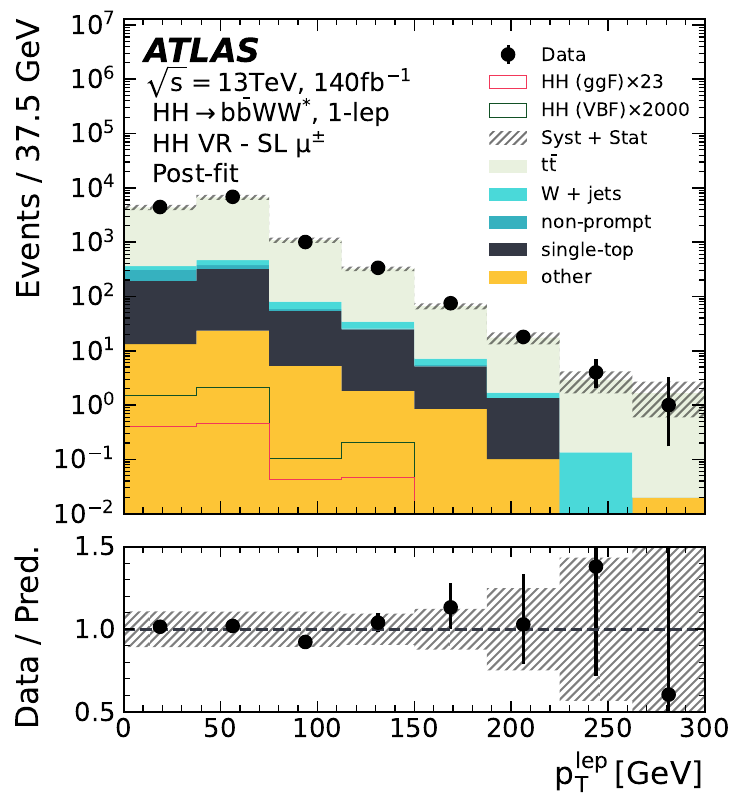}}
\subfloat[]{\includegraphics[width=0.45\textwidth]{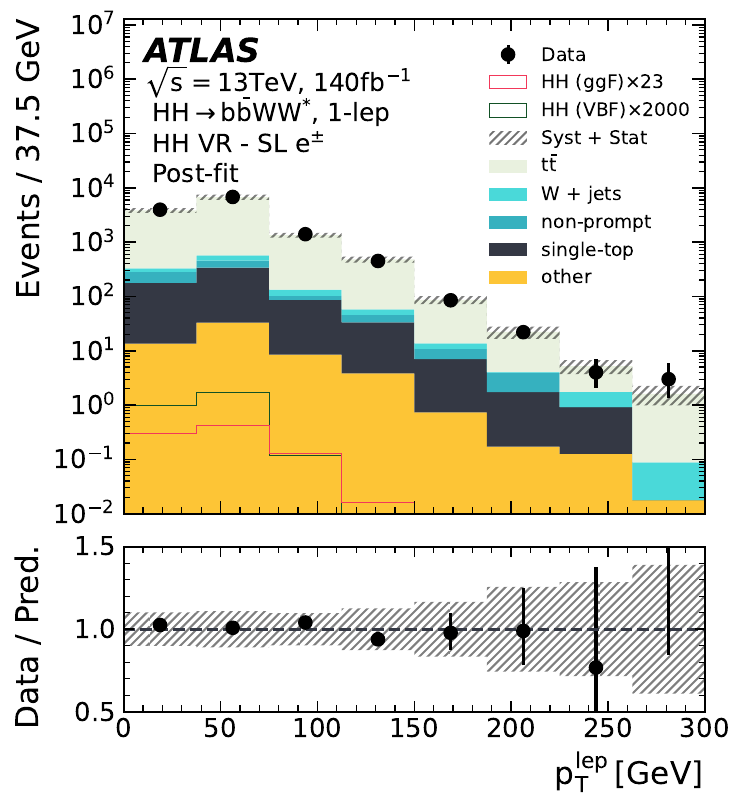}}
\end{center}
\caption{Distribution of the \pt of the (a) electrons and (b) muons in the HH VR. The
hashed band includes all systematic uncertainties considered in the analysis. Backgrounds are normalised using post-fit normalisation factors.} \label{fig:KF_inputs_validation}
\end{figure}

The ability of the Monte Carlo simulation to reproduce the shape of the input variables is tested in a dedicated validation region (HH VR), defined by selecting events with log$(T) \le 4$ and a cut on the Neural Network discriminant (\Sect{\ref{sec:NN}}) $d_{HH} > -3$,
the results obtained after fitting the signal and the background normalisation factors (\Sect{\ref{sec:MainResults}}) are shown in \Fig{\ref{fig:KF_inputs_validation}} for the lepton \pt distributions as an example.


\section{Systematic Uncertainties}
\label{sec:systematics}

Various sources of systematic uncertainties affect the analysis.
They consist in experimental uncertainties, from both detector and
reconstruction effects, and in theoretical uncertainties
on the modelling of the physics processes.
All uncertainties are included in the likelihood fit as nuisance parameters (NP) as they can impact the signal and background yield,
and the shape of the fitted distributions.

Uncertainties are evaluated on lepton ($e$,$\mu$,$\tau_{\mathrm{had}}$) reconstruction, identification and momentum/energy
calibration~\cite{EGAM-2018-01,EGAM-2021-01,EGAM-2021-02,MUON-2018-03,MUON-2022-01,ATL-PHYS-PUB-2019-033},
\met reconstruction~\cite{JETM-2020-03}, jet energy scale, resolution and vertex tagging~\cite{JETM-2018-05,ATL-PHYS-PUB-2019-026},
$b$-jet identification~\cite{FTAG-2019-07}, large-$R$ jet energy scale, resolution, and mass scale~\cite{JETM-2024-03}.
An uncertainty of 0.83\%~\cite{DAPR-2021-01}
is applied to the expected yields of the simulated samples to account for the uncertainty on the integrated
luminosity as measured by the LUCID-2 detector~\cite{LUCID2}.
All of these experimental uncertainties are treated as correlated among the signal and background processes, and are found to have a negligible impact on the result.

Modelling uncertainties associated with the $t\bar{t}$, $W$+jets, and single-top processes are evaluated by comparing the nominal MC distributions
with those from the alternative MC generator predictions described in \Sect{\ref{sec:samples}} in each SR and CR.
Additional uncertainties due to missing higher order corrections are evaluated from variations of the renormalisation and factorisation scales in the matrix element
calculations by a factor of 0.5 and 2. For $t\bar{t}$, single-top and the $HH$ signal, uncertainties on the initial-state and final-state radiation simulation are computed by varying the MC tuning computed with $\alpha_s$ values corresponding to upward and downward variations of the renormalisation scale.
The PDF uncertainties are computed using the eigenvectors of the PDF set used depending on the sample, and following the prescriptions in Ref.~\cite{Butterworth:2015oua,Ball:2014uwa}.
Uncertainties on the theoretical production cross-sections of all the simulated processes are also taken into account as nuisance parameters.

Three sources of uncertainty are identified in the estimation of data-driven fakes.
The variation of the fake-factor due to $t\bar{t}$ normalisation uncertainty is determined from the difference
between the measured normalisation and its theoretical prediction.
The extrapolation uncertainty accounts for differences in background composition between the fakes-enriched control region and the signal region.
These differences arise in the relative fractions of $b$-hadron decays compared to $c$-hadron and light-flavour decays.
Corresponding fake-factor uncertainties are extracted from the MC simulation by comparing the estimated yield obtained applying the fake-factor method to the MC sample and the expected yield from MC itself.
The fake-factor statistical uncertainty includes both data and MC statistical uncertainty in the fakes-enriched region.

Systematic uncertainties in the modelling of the non-resonant signal are computed by varying the renormalisation and factorisation scales with a variation of up to a factor of two,
and using the same procedure adopted for the $t\bar{t}$, $W$+jets, and single-top backgrounds.
PDF uncertainties are computed using PDF4LHC15\_30~\cite{Butterworth:2015oua} PDF sets, which include the envelope of three PDF sets, namely CT14, MMHT14 and NNPDF3.0.
For the non-resonant signal, parton shower uncertainties are also computed by comparing the \textsc{Herwig++} showering with that of \textsc{Pythia8}.

The impact of various categories of uncertainties on the observed $\sigma/\sigma_{\textrm{exp}}$ parameter (where $\sigma$ is the measured cross-section and $\sigma_{\textrm{exp}}$ is the cross-section used to normalise the signal MC sample) is quantified. $\sigma/\sigma_{\textrm{exp}}$  is implemented in the fit model as a
multiplicative factor applied to the expected yield for the ggF+VBF $HH$ and the $X \to SH$ signal.
Table~\ref{tab:sysimpact} summarises the impact on the $\sigma/\sigma_{\textrm{exp}}$ uncertainty for the non-resonant $HH$ signal, and for the $X \to SH$ signal with the mass values $m_X=750$~\GeV, $m_S=240$~\GeV and $m_X=3000$~\GeV, $m_S=2500$~\GeV.
The impact of each NP on $\sigma/\sigma_{\textrm{exp}}$ is evaluated by fixing the considered
NP to its best-fit value and subtracting the obtained uncertainty in quadrature from the total uncertainty.
The differences in the relative contributions of modelling uncertainties in the different SRs arise from the difference in the background compositions.

\begin{table}
\caption{Contribution of systematic uncertainties to the total uncertainty on the observed $\sigma/\sigma_{
\textrm{exp}}$.
The impact of each parameter is normalised to the total uncertainty on $\sigma/\sigma_{\textrm{exp}}$. For comparison, the statistical uncertainties from data and MC are also shown.
The ``Other'' systematic uncertainties consist of the electron and muon identification, reconstruction, isolation and trigger efficiencies; the $\tau_{\mathrm{had}}$ identification and reconstruction efficiencies; the cross-sections of the subdominant backgrounds; the JVT efficiency; the pile-up reweighting; and the luminosity.}
\label{tab:sysimpact}
\begin{center}
\begin{tabular}{l|c|c|c}
\toprule
Uncertainty source & \multicolumn{3}{c}{Fractional contribution [\%]} \\
& non-resonant $HH$ & \multicolumn{2}{c}{$X \to SH$} \\
&  & $m_X = 750$~\GeV & $m_X = 3000$~\GeV \\
&  & $m_S = 240$~\GeV & $m_S = 2500$~\GeV \\
\midrule
\textbf{Systematic uncertainties}            &    &    &    \\
Signal modelling                             & 3  & -  & -  \\
$t\bar{t}$ modelling                         & 48 & 28 & 46 \\
$W$+jets modelling                           & 2  & 5  & 38 \\
single-top modelling                         & 18 & 16 & 48 \\
Fakes modelling                              & 31 & 19 & 10 \\
Fakes statistics                             & 33 & 46 & 11 \\
Flavour tagging                              & 25 & 10 & 6  \\
Small-$R$ jet energy scale and resolution    & 14 & 26 & 36 \\
Large-$R$ jet energy scale and resolution    & -  & 4  & 4  \\
Other                                        & 9  & 9  & 12 \\
MC stat. uncertainty                         & 39 & 23 & 22 \\
\midrule
\textbf{Total syst. uncertainty}             & 84 & 78 & 77 \\
\midrule
\textbf{Data stat. uncertainty}              & 54 & 63 & 64 \\
\bottomrule
\end{tabular}
\end{center}
\end{table}


%
\section{Results}
\label{sec:MainResults}

A maximum-likelihood fit is performed for each signal hypothesis on all bins in the 6 CRs + 4 SRs event categories defined in \Sect{\ref{sec:selection}}
for the non-resonant $HH$. For the resonant $X \to SH$ analysis an independent  fit is performed for each ($m_X, m_S$) hypothesis using the corresponding CRs and SRs. The fit determines the normalisation factors of the major backgrounds
and upper limits on the signal cross-section.
Inclusive normalisation factors are used for the $t\bar{t}$ and $W$+jets processes, while
two normalisation factors are used for the single-top process: one for event triggered by SL triggers and one for  MET triggered events. In addition
two independent normalisation factors are used for the multijet background events depending on the selected lepton flavour ($e$/$\mu$). Uncertainties affecting the total yield of the normalised backgrounds are included in the fit, although their impacts are effectively cancelled in the fit model.

The likelihood function $\mathcal{L}(\sigma/\sigma_{\textrm{exp}},\lambda,\theta)$ is constructed as a product of Poisson probability terms, $\mathrm{Pois}(N_{\mathrm{obs}}|\sigma/\sigma_{\textrm{exp}}\cdot s+b)$,
for observing $N_{\mathrm{obs}}$ events given $\sigma/\sigma_{\textrm{exp}}\cdot s$ signal and $b$ background events in each bin of the fitted distribution considered in the analysis.
It depends on the signal multiplying factor $\sigma/\sigma_{\textrm{exp}}$, on the background normalisation factors $\lambda$, and on the NPs encoding the systematic uncertainties
in the signal and background expectations.
The total event yield (i.e. a single bin) is used in each control region, while the $d_{HH}$ (or $d_{SHi}$) output DNN distribution
is employed as final discriminant in the signal regions defined in Figure~\ref{fig:regions} and Table~\ref{tab:sh_srvr_binning}.
The statistical model is implemented using HistFactory~\cite{Cranmer:2012sba,RooFit,Moneta:2010pm}.

\begin{figure}
\begin{center}
\subfloat[]{
\includegraphics[width=0.45\textwidth]{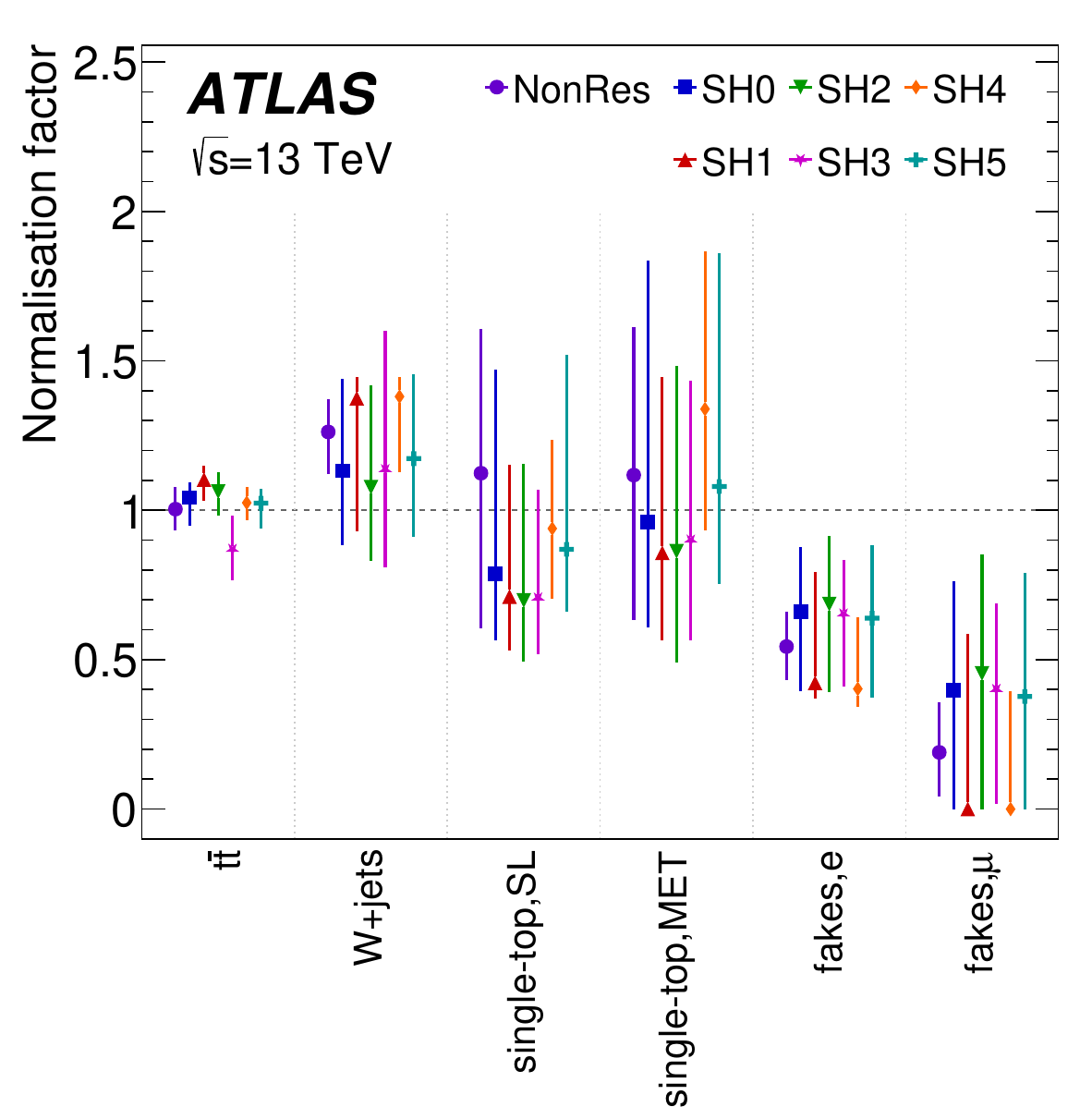}
}
\subfloat[]{
\includegraphics[width=0.45\textwidth]{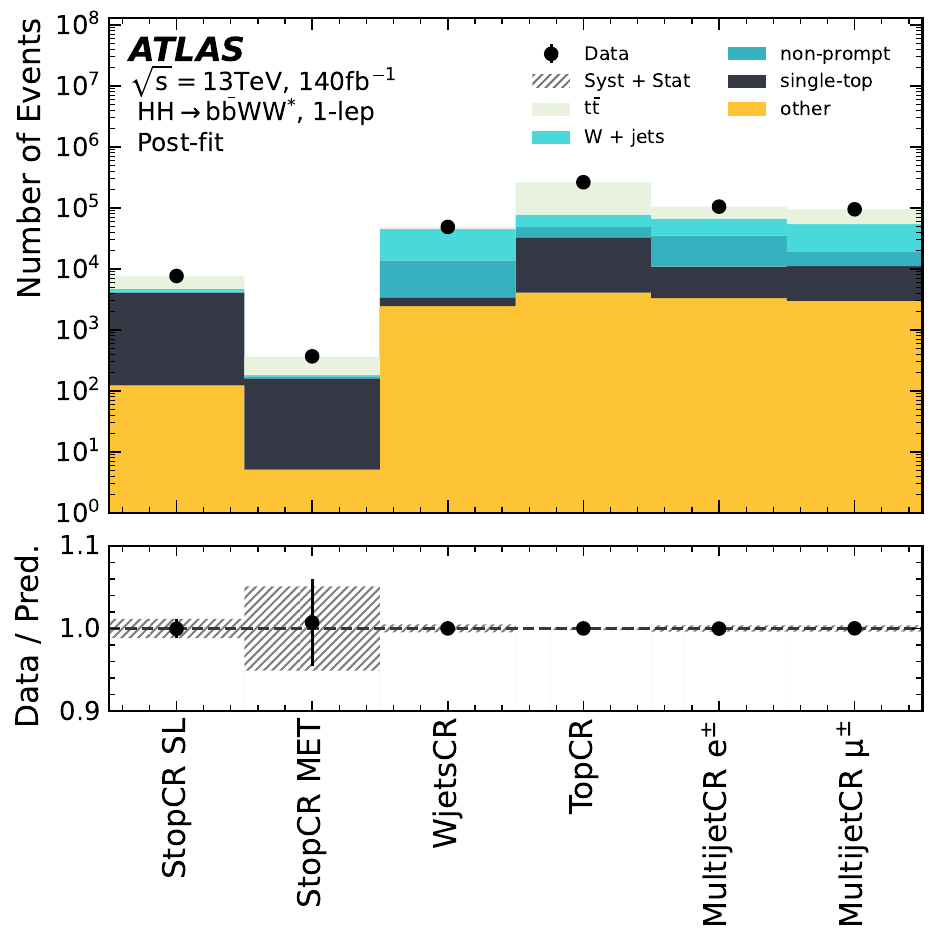}
}
\end{center}
\caption{(a) Post-fit normalisation factors for the main backgrounds in the non-resonant and $SH$ analysis for CR + SR fits. The error bars represents statistical and systematic uncertainty.
(b) Post-Fit comparison between data and the background prediction for the event yields in the six non-resonant analysis CRs. The ``other'' background include $Z$+jets, diboson, $ttV$ and single-Higgs processes.
The lower panel shows the ratio of data over the total expected background (``Pred.'') with the corresponding statistical and total uncertainty.
The size of the combined statistical and systematic uncertainty in the background prediction is indicated by the hatched band.
}
\label{fig:bkg_CRs}
\end{figure}

Background normalisation factors and NP values are derived from the fit. Post-fit background normalisation factors and event yields in the six $HH$ analysis CRs are presented in \Fig{\ref{fig:bkg_CRs}}. The fake factor method is used to determine the multijet acceptance in the SRs and CRs, alongside the shape of the NN discriminant distributions. The multijet normalisation is determined from the fit to data; however, the uncertainties shown in the plot do not include the large global normalisation component from fake factor systematic uncertainties, as this source is cancelled by the fit model and not accounted for in the error bars.

\begin{figure}
\begin{center}
\subfloat[]{
\includegraphics[width=0.4\textwidth]{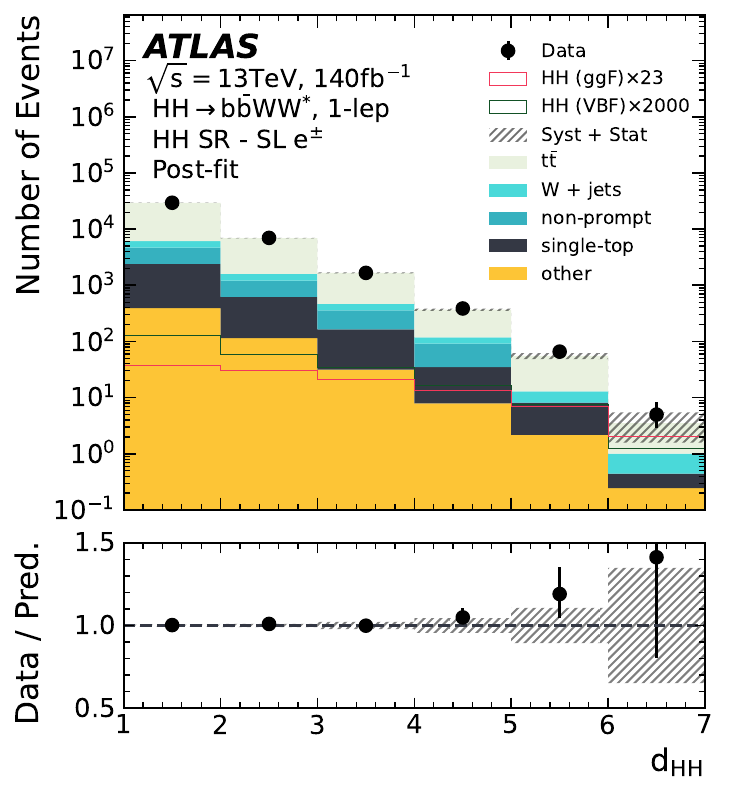}
\label{fig:dHH_eslSR}
}
\subfloat[]{
\includegraphics[width=0.4\textwidth]{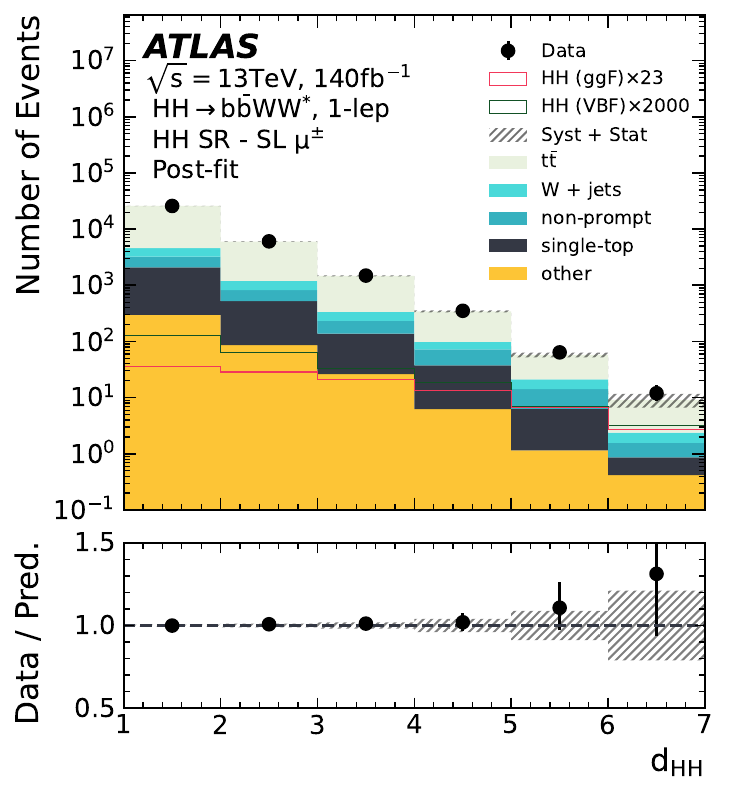}
\label{fig:dHH_mslSR}
}\\
\subfloat[]{
\includegraphics[width=0.4\textwidth]{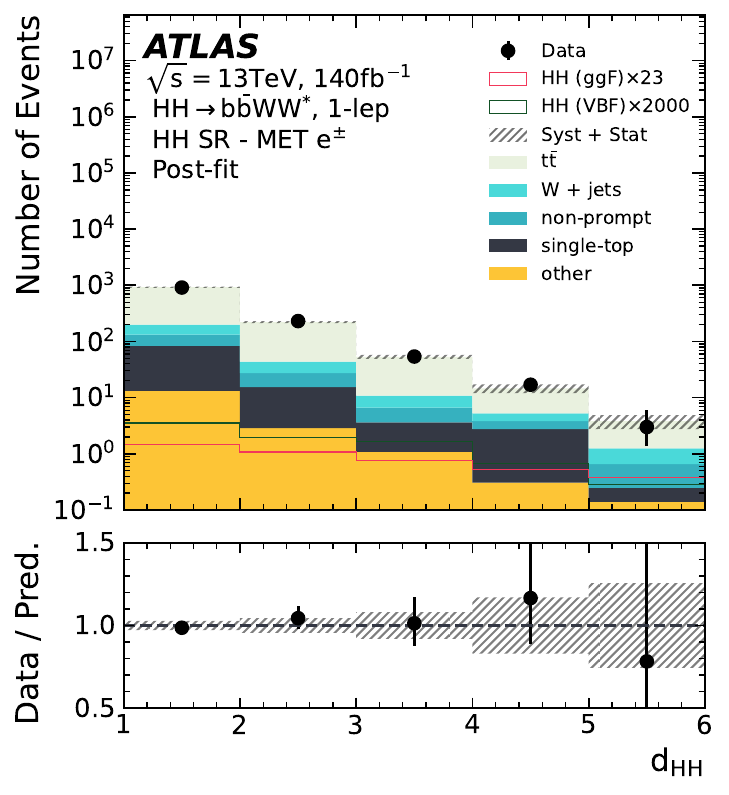}
\label{fig:dHH_emetSR}
}
\subfloat[]{
\includegraphics[width=0.4\textwidth]{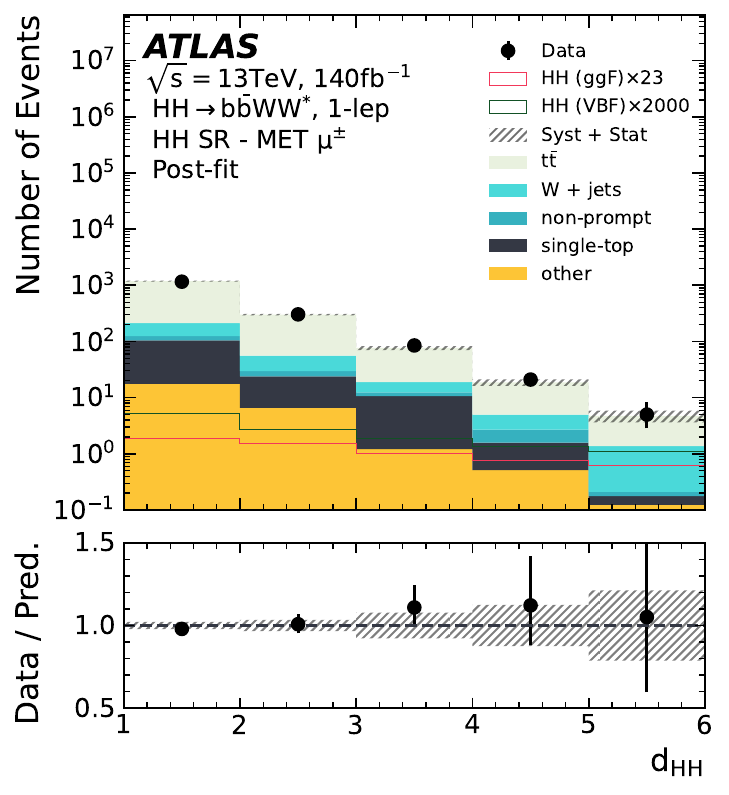}
\label{fig:dHH_mmetSR}
}
\end{center}
\caption{Comparison between data and the background prediction in the $d_{HH}$ distributions of the non-resonant analysis,
for the event yields in the four non-resonant analysis SRs (HH SR): (a) SL $e^{\pm}$, (b) SL $\mu^{\pm}$, (c) MET $e^{\pm}$, (d) MET $\mu^{\pm}$.
The last bin includes the overflow events.
For each SR the non-resonant ggF and VBF $HH$ signal yields are shown, scaled by their expected upper-limit cross-section reported in this Section. The ``other'' background include $Z$+jets, diboson, $ttV$ and single-Higgs processes.
The background contributions after the likelihood fit to data (``Post-Fit'') under the background-only hypothesis are shown as filled histograms.
The lower panel shows the ratio of data divided by the total expected background (``Pred.'') with the corresponding statistical and total uncertainty.
The size of the combined statistical and systematic uncertainty in the background prediction is indicated by the hatched band.}
\label{fig:dHH_SRs}
\end{figure}

Good agreement of data with the MC expectation is observed.
The post-fit distributions for the prediction compared with the data in
the SRs as a function of $d_{HH}$, the fit discriminant,
are shown in \figurename~\ref{fig:dHH_SRs}.

Small excesses are observed in moderately high $s/b$ regions in all
channels. However the background predictions uncertainties are correlated between
them, and the combined significance is only 1.1 sigma.

The results are used to evaluate an upper limit at the $95\%$ confidence level (CL) on the production cross-section times
the branching fraction for the signal hypotheses under consideration.
The exclusion limits are calculated with the CL$_{\mathrm{s}}$ method~\cite{CLs_2002} and the
profile-likelihood test statistic~\cite{asymptotics}.
Among the considered systematic uncertainties, only background modelling NPs are slightly constrained, with the most significant
constraint being a factor two on the single top DS/DR systematic, while the detector related uncertainties
are not significantly constrained or pulled in the likelihood fit.

With the non-resonant signal hypothesis, the observed (expected) upper limit at 95\% CL is:
\[
\sigma(pp \to HH) \cdot{\mathcal{B}}(HH \to bbWW^{\ast}) < 268 \, \left (178^{+78}_{-54} \right )  \,{\mathrm{fb}}.
\]
The branching fraction ${\mathcal{B}}(HH \to bbWW^{\ast}) = 2\times{\mathcal{B}}(H \to bb) \times {\mathcal{B}}(H \to WW^{\ast}) = 0.249$ is used to
obtain the following observed (expected) limit on the ggF $HH$ production cross-section at 95\% CL:
\[
\sigma(pp \to HH) < 1.08 \, \left (0.72^{+0.31}_{-0.22} \right) \,  {\mathrm{pb}},
\]
which corresponds to $35\,(23^{+10}_{-7})$ times the SM predicted cross-section, with no excess over the SM prediction found.
These results do not take into account the SM $HH$ cross-section uncertainty.
Including only the statistical uncertainty, the expected upper limit for the non-resonant production is 16.2 times the SM prediction.
The previous ATLAS result in the same channel obtained with $36.1\,\mathrm{fb}^{-1}$~\cite{HIGG-2016-27}, set an observed (expected) limit as $300\,(300^{+100}_{-80})$ times the SM predicted cross-section. The expected sensitivity of the new result therefore improves by a factor 13, of which a factor 2 can be attributed to the larger data sample,
while the remaining factor of about 6.6 is obtained thanks to the improvements of the analysis method.

The analysis sensitivity is similar (slightly better) than the one performed by the CMS Collaboration for the same $HH\to bbWW$ single-lepton decay channel~\cite{CMS-HIG-21-005},
but the final result shows a slight over-fluctuation of the observed upper limit on the signal cross-section.
While the analysis is not competitive with other $HH$ decay channels mainly due to
the large and irreducible $t\bar{t}$ background, the use of Machine Learning techniques in the
background discrimination has significantly improved the previous ATLAS search.

 %

\begin{figure}
\begin{center}
\includegraphics[width=0.99\textwidth]{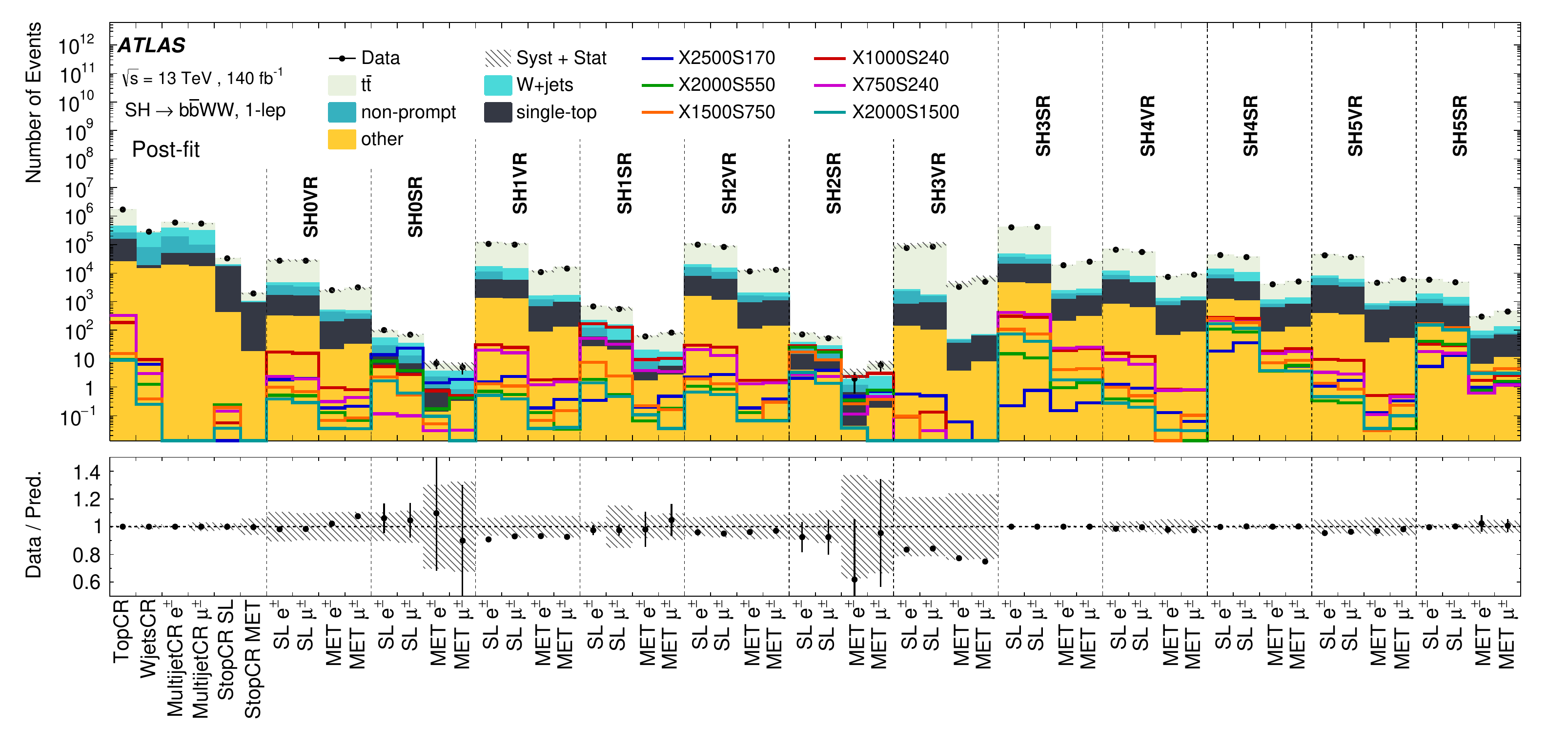}
\label{fig:SH_yields_postfit}
\end{center}
\caption{Post-Fit comparison between data and the background prediction for the event yields in the $SH$ CRs, VRs and SRs.
The lower panel shows the ratio of data over the total expected background prediction (``Pred.'').
The size of the combined statistical and systematic uncertainty in the
background prediction is indicated by the hatched band.
Six signal mass points targeted by different SRs are shown for reference.
A cross-section of 100 fb is used to normalise the signal predictions.}
\label{fig:SH_yields}
\end{figure}

The agreement between data and simulation is also validated
in the $SH$ analysis for all the CRs and VRs.
In all validation regions, data are in agreement with the expectation.
In order to reduce strongly constrained  systematic uncertainties in the SH3VR validation region, due to the large statistics of the corresponding SR, the major theoretical uncertainties on the $t\bar{t}$ background were decorrelated among signal, control and validation regions. This reduces the constraints at the cost of increasing the systematic error in those regions. The systematic uncertainties affected are  fully correlated among all bins of the region. No statistically significant discrepancy from the SM prediction is observed in any of the $SH$ SRs, see \Fig{\ref{fig:SH_yields}}.

\begin{figure}[!h]
\begin{center}
\includegraphics[width=0.9\textwidth,clip]{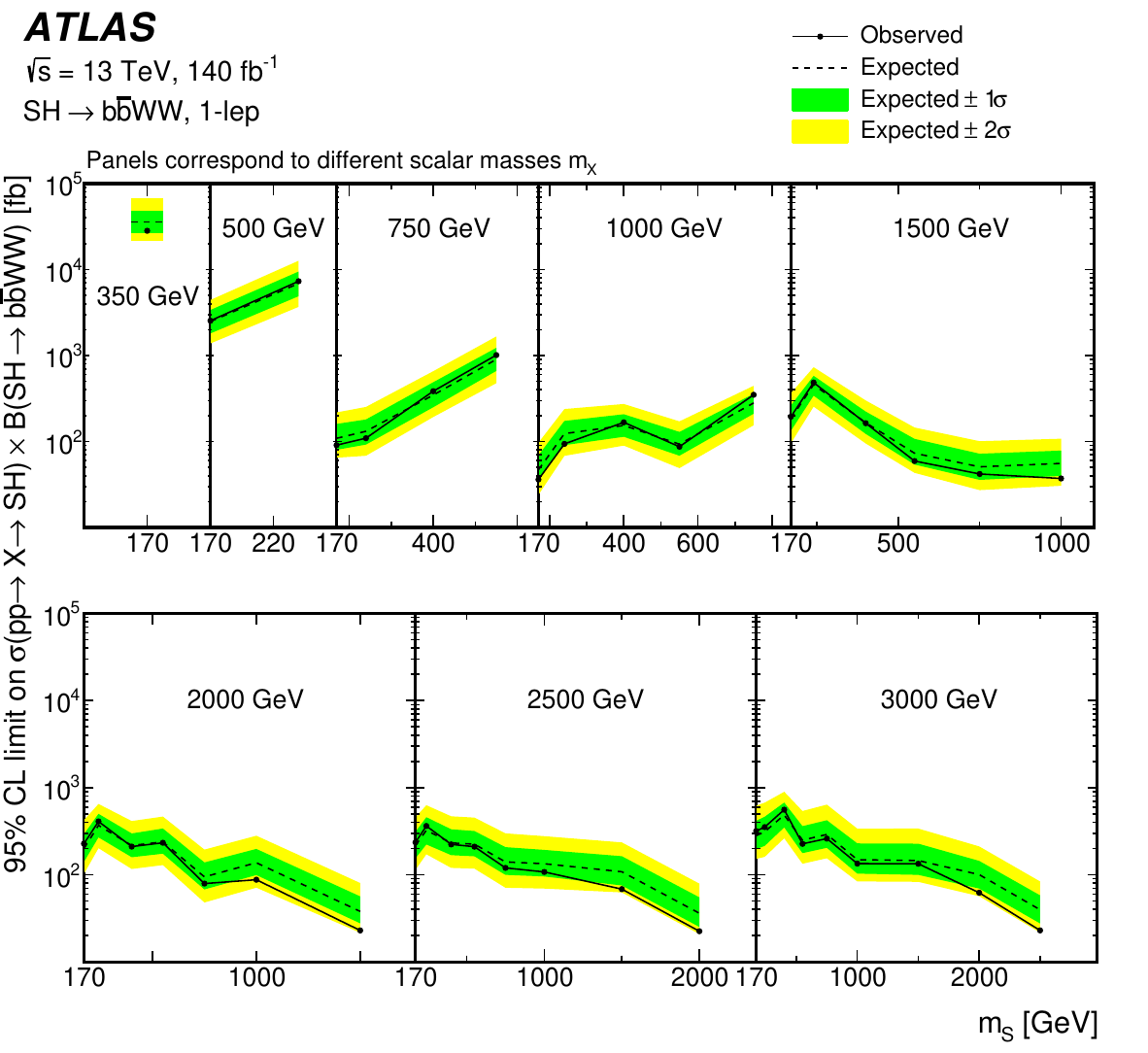}\\
\end{center}
\caption{Expected and observed upper limits at 95\% CL on $\sigma(pp\to X)\times\mathcal{B}(X\to SH)\times\mathcal{B}(SH\to b\bar{b} WW)$
as a function of $m_S$ for each value of the $X$ scalar mass $m_X$.} \label{fig:SHlimit_summary}
\end{figure}

The $t\bar{t}$ uncertainites are generally constrained by the large number of events in data and the significant $t\bar{t}$ contribution to the SRs. For example, in SR3 the conservative $t\bar{t}$ uncertainty estimated by scale variations at NLO is reduced by a factor of about three.
Exclusion limits at 95\% CL$_{\mathrm{s}}$ on the $pp \to X \to SH \to b\bar{b}WW$ production cross-section multiplied by the branching fraction, denoted as
$\sigma$($pp \to X \to SH) \times \mathcal{B} (SH \to b\bar{b}WW$), are determined for several $m_X$ and $m_S$ hypotheses.
The expected and observed limits are shown in \Fig{\ref{fig:SHlimit_summary}} for each combination of
the scalar masses.
While limits can be safely interpolated along the $m_S$ variable,  they are valid only at the chosen $m_X$ mass values.

          %

%

%
%
%

%
\section{Conclusions}
\label{sec:conclusion}

A search for $b\bar{b}W^+W^-$ events produced from the decay of non-resonant Higgs boson pairs or produced through the resonant $X\to SH$ process
is presented. The analysis is performed using $pp$ collision data corresponding to an integrated luminosity of 140~fb$^{-1}$, collected at $\sqrt{s}=13$~\TeV by the
ATLAS detector at the Large Hadron Collider. Final states with one lepton (electrons, muons or leptonically decaying $\tau$-leptons) and resolved $H \to b\bar{b}$ and $W \to jj$ decays are studied.
In the non-resonant analysis, the SM $HH$ production is investigated, while in the $SH$ analysis
several combinations of the
two scalar's masses within the $m_X$ mass range $350-3000$~\GeV and $m_S$ mass range $170-2500$~\GeV are probed.

In the non-resonant analysis, no significant excess of events over the background expectation is found.
The observed (expected) limit on the non-resonant Higgs pair production cross-section ratio $\sigma_{HH}/\sigma_{HH}^{SM}$ is 35 ($23^{+10}_{-7}$),
the observed limit is within $+1.1\sigma$ from its expectation value.
In the $X \to SH$ signal hypothesis, no signal is found over the expected background distribution, and exclusion limits on the cross-section
$\sigma(pp\to X\to SH) \times \mathcal{B}(SH\to b\bar{b}WW)$ are set.
Systematic uncertainties play an important role, yet the results show a large improvement over the previous ATLAS results in the same $HH$ decay channel:
the limits on non-resonant $HH$ production are stronger by a factor of 13, where a factor of two can be derived from the enlarged data sample and the remaining factor of almost seven is obtained from optimised analysis procedures.
The use of a Neural Network discriminant, an improved multijet background rejection and estimate, and a better design of the background CRs are the main contributors to this improvement. Furthermore, this work provides a new result on the resonant $pp\to X\to SH$, never investigated so far in the $bbWW$ 1-lepton decay channel.

          %
%

%
\section*{Acknowledgements}
%

%
%

%
%

We thank CERN for the very successful operation of the LHC and its injectors, as well as the support staff at
CERN and at our institutions worldwide without whom ATLAS could not be operated efficiently.

The crucial computing support from all WLCG partners is acknowledged gratefully, in particular from CERN, the ATLAS Tier-1 facilities at TRIUMF/SFU (Canada), NDGF (Denmark, Norway, Sweden), CC-IN2P3 (France), KIT/GridKA (Germany), INFN-CNAF (Italy), NL-T1 (Netherlands), PIC (Spain), RAL (UK) and BNL (USA), the Tier-2 facilities worldwide and large non-WLCG resource providers. Major contributors of computing resources are listed in Ref.~\cite{ATL-SOFT-PUB-2026-001}.

We gratefully acknowledge the support of ANPCyT, Argentina; YerPhI, Armenia; ARC, Australia; BMWFW and FWF, Austria; ANAS, Azerbaijan; CNPq and FAPESP, Brazil; NSERC, NRC and CFI, Canada; CERN; ANID, Chile; CAS, MOST, NSFC and FRFCU, China; Minciencias, Colombia; MEYS CR, Czech Republic; DNRF and DNSRC, Denmark; IN2P3-CNRS and CEA-DRF/IRFU, France; SRNSFG, Georgia; BMFTR, HGF and MPG, Germany; GSRI, Greece; RGC and Hong Kong SAR, China; ICHEP and Academy of Sciences and Humanities, Israel; INFN, Italy; MEXT and JSPS, Japan; CNRST, Morocco; NWO, Netherlands; RCN, Norway; MNiSW, Poland; FCT, Portugal; MNE/IFA, Romania; MSTDI, Serbia; MSSR, Slovakia; ARIS and MVZI, Slovenia; DSI/NRF, South Africa; MICIU/AEI, Spain; SRC and Wallenberg Foundation, Sweden; SERI, SNSF and Cantons of Bern and Geneva, Switzerland; NSTC, Taipei; TENMAK, T\"urkiye; STFC/UKRI, United Kingdom; DOE and NSF, United States of America.

Individual groups and members have received support from BCKDF, CANARIE, CRC and DRAC, Canada; CERN-CZ, FORTE and PRIMUS, Czech Republic; COST, ERC, ERDF, Horizon 2020 and Marie Sk{\l}odowska-Curie Actions, European Union; Investissements d'Avenir Labex, Investissements d'Avenir Idex and ANR, France; DFG and AvH Foundation, Germany; Herakleitos, Thales and Aristeia programmes co-financed by EU-ESF and the Greek NSRF, Greece; BSF-NSF and MINERVA, Israel; NCN and NAWA, Poland; La Caixa Banking Foundation, CERCA and AGAUR programs from Generalitat de Catalunya and PROMETEO and GenT Programmes Generalitat Valenciana, Spain; G\"{o}ran Gustafssons Stiftelse, Sweden; The Royal Society and Leverhulme Trust, United Kingdom; Eric and Wendy Schmidt Fund for Strategic Innovation, United States of America.

In addition, individual members wish to acknowledge support from Chile: Agencia Nacional de Investigaci\'on y Desarrollo (ANID FONDECYT reg. 1230987, FONDECYT 1230812, FONDECYT 1240864, Fondecyt 3240661, Fondecyt Regular 1240721); China: Fundamental Research Funds for the Central Universities (010-63263105), Chinese Ministry of Science and Technology (MOST-2023YFA1605700, MOST-2023YFA1609300), National Natural Science Foundation of China (NSFC 12275265, NSFC-W2543005); Czech Republic: Czech Science Foundation (GACR - 24-11373S), Ministry of Education Youth and Sports (ERC-CZ-LL2327, FORTE CZ.02.01.01/00/22\_008/0004632); EU: H2020 European Research Council (ERC - 101002463); European Union: European Research Council (BARD No. 101116429, ERC - 101219398, ERC 101089007), European Regional Development Fund (HE COFUND GA No.101081355, ERDF), Marie Sklodowska-Curie Actions (GAP-101168829); France: Agence Nationale de la Recherche (ANR-22-EDIR-0002, ANR-24-CE31-0504-01); Germany: Deutsche Forschungsgemeinschaft (DFG - 469666862); China: Research Grants Council (GRF); Italy: Istituto Nazionale di Fisica Nucleare (LHC-MIUR - 28003/2025), Ministero dell'Università e della Ricerca (NextGenEU I53D23000820006 M4C2.1.1, SOE2024\_0000023); Japan: Japan Society for the Promotion of Science (JSPS KAKENHI  JP25H0063, JSPS KAKENHI JP22H04944, JSPS KAKENHI JP24K23939, JSPS KAKENHI JP24KK0251, JSPS KAKENHI JP25H00650, JSPS KAKENHI JP25H01291, JSPS KAKENHI JP25K01011, JSPS KAKENHI JP25K01023, JSPS KAKENHI JP25KK0047); Poland: Polish National Science Centre (NCN OPUS 2023/51/B/ST2/02507, NCN OPUS nr 2022/47/B/ST2/03059, UMO-2019/34/E/ST2/00393, UMO-2021/42/E/ST2/00350, UMO-2022/47/O/ST2/00148, UMO-2023/49/B/ST2/04085, UMO-2023/51/B/ST2/00920, UMO-2024/53/N/ST2/00869); Spain: Agència de Gestió d'Ajuts Universitaris i de Recerca. (AGAUR - 2023 BP 00141), Ministry of Science and Innovation (RYC2019-028510-I, RYC2020-030254-I, RYC2021-031273-I, RYC2022-038164-I), Ministerio de Ciencia, Innovación y Universidades/Agencia Estatal de Investigaci\'on (EU NextGenerationEU (PRTR-C17.I1), PID2022-142604OB-C22); Sweden: Carl Trygger Foundation (Carl Trygger Foundation CTS 22:2312), Swedish Research Council (Swedish Research Council 2023-04654, VR 2021-03651, VR 2022-03845, VR 2022-04683, VR 2023-03403, VR 2024-05451, VR 2025-05940), Knut and Alice Wallenberg Foundation (KAW 2023.0366); United Kingdom: The Binks Trust, Royal Society (NIF-R1-231091); United States of America: U.S. Department of Energy (ECA DE-AC02-76SF00515), John Templeton Foundation (John Templeton Foundation 63206), Neubauer Family Foundation.

%
%


%
%
%
%
%
%
%

%
%
%
%
%
%

%
%
%
\addcontentsline{toc}{part}{References}
\printbibliography
\clearpage
 
\begin{flushleft}
\hypersetup{urlcolor=black}
{\Large The ATLAS Collaboration}

\bigskip

\AtlasOrcid[0000-0002-6665-4934]{G.~Aad}$^\textrm{\scriptsize 102}$,
\AtlasOrcid[0000-0001-7616-1554]{E.~Aakvaag}$^\textrm{\scriptsize 17}$,
\AtlasOrcid[0000-0002-5888-2734]{B.~Abbott}$^\textrm{\scriptsize 121}$,
\AtlasOrcid[0000-0002-0287-5869]{S.~Abdelhameed}$^\textrm{\scriptsize 83b}$,
\AtlasOrcid[0000-0002-1002-1652]{K.~Abeling}$^\textrm{\scriptsize 54}$,
\AtlasOrcid[0000-0001-5763-2760]{N.J.~Abicht}$^\textrm{\scriptsize 48}$,
\AtlasOrcid[0000-0002-8496-9294]{S.H.~Abidi}$^\textrm{\scriptsize 30}$,
\AtlasOrcid[0009-0003-6578-220X]{M.~Aboelela}$^\textrm{\scriptsize 44}$,
\AtlasOrcid[0000-0002-9987-2292]{A.~Aboulhorma}$^\textrm{\scriptsize 36e}$,
\AtlasOrcid[0000-0001-5329-6640]{H.~Abramowicz}$^\textrm{\scriptsize 154}$
% ...
\bigskip
\\

% ...
$^{17}$Department for Physics and Technology, University of Bergen, Bergen; Norway.\\
% ...
$^{30}$Physics Department, Brookhaven National Laboratory, Upton NY; United States of America.\\
% ...
$^{36}$$^{(a)}$Facult\'e des Sciences Ain Chock, Universit\'e Hassan II de Casablanca;$^{(b)}$Facult\'{e} des Sciences, Universit\'{e} Ibn-Tofail, K\'{e}nitra;$^{(c)}$Facult\'e des Sciences Semlalia, Universit\'e Cadi Ayyad, LPHEA-Marrakech;$^{(d)}$LPMR, Facult\'e des Sciences, Universit\'e Mohamed Premier, Oujda;$^{(e)}$Facult\'e des sciences, Universit\'e Mohammed V, Rabat;$^{(f)}$Institute of Applied Physics, Mohammed VI Polytechnic University, Ben Guerir; Morocco.\\
% ...
$^{44}$Physics Department, Southern Methodist University, Dallas TX; United States of America.\\
% ...
$^{48}$Fakult\"{a}t Physik, Technische Universit{\"a}t Dortmund, Dortmund; Germany.\\
% ...
$^{54}$II. Physikalisches Institut, Georg-August-Universit\"{a}t G\"ottingen, G\"ottingen; Germany.\\
% ...
$^{83}$$^{(a)}$Khalifa University of Science and Technology, Abu Dhabi;$^{(b)}$New York University Abu Dhabi, Abu Dhabi;$^{(c)}$United Arab Emirates University, Al Ain;$^{(d)}$University of Sharjah, Sharjah; United Arab Emirates.\\
% ...
$^{102}$CPPM, Aix-Marseille Universit\'e, CNRS/IN2P3, Marseille; France.\\
% ...
$^{121}$Homer L. Dodge Department of Physics and Astronomy, University of Oklahoma, Norman OK; United States of America.\\
% ...
$^{154}$Raymond and Beverly Sackler School of Physics and Astronomy, Tel Aviv University, Tel Aviv; Israel.\\
% ...
$^{a}$ Also at Affiliated with an institute formerly covered by a cooperation agreement with CERN.\\
$^{b}$ Also at An-Najah National University, Nablus; Palestine.\\
$^{c}$ Also at Borough of Manhattan Community College, City University of New York, New York NY; United States of America.\\
$^{d}$ Also at Center for Interdisciplinary Research and Innovation (CIRI-AUTH), Thessaloniki; Greece.\\
$^{e}$ Also at Centre of Physics of the Universities of Minho and Porto (CF-UM-UP); Portugal.\\
$^{f}$ Also at CERN, Geneva; Switzerland.\\
% ...

\end{flushleft}


%

%
%

%

%
%
%
%

\end{document}